\pdfoutput=1 
\PassOptionsToPackage{table}{xcolor} 
\documentclass[sigconf]{acmart}
\usepackage{amsmath} 
\usepackage{amsfonts} 
\usepackage{booktabs} 
\usepackage{multirow} 
\usepackage{makecell} 
\usepackage{graphicx} 
\usepackage{subcaption} 
\usepackage[table]{xcolor} 
\usepackage{tcolorbox} 
\usepackage{CJKutf8} 
\usepackage[justification=centering]{caption} 
\usepackage{enumitem} 
\usepackage{float} 
\usepackage[ruled,linesnumbered]{algorithm2e} 
\usepackage{hyperref} 
\hypersetup{
  hidelinks, 
  colorlinks=true,
  linkcolor=blue,
  citecolor=red
}
\usepackage{balance}

\copyrightyear{2025}
\acmYear{2025}
\setcopyright{acmlicensed}\acmConference[KDD '25]{Proceedings of the 31st ACM SIGKDD Conference on Knowledge Discovery and Data Mining V.2}{August 3--7, 2025}{Toronto, ON, Canada}
\acmBooktitle{Proceedings of the 31st ACM SIGKDD Conference on Knowledge Discovery and Data Mining V.2 (KDD '25), August 3--7, 2025, Toronto, ON, Canada}
\acmDOI{10.1145/3711896.3737106}
\acmISBN{979-8-4007-1454-2/2025/08}

\begin{document}

\title[Quadratic Neural Networks for Click-through Rate Prediction]{Revisiting Feature Interactions from the Perspective of Quadratic Neural Networks for Click-through Rate Prediction}

\author{Honghao Li}
\email{salmon1802li@gmail.com}
\orcid{0009-0000-6818-7834}
\affiliation{%
  \institution{Anhui University}
  \city{Hefei}
  \state{Anhui Province}
  \country{China}
}

\author{Yiwen Zhang}
\authornote{Yiwen Zhang (zhangyiwen@ahu.edu.cn) is the Corresponding author.}
\orcid{0000-0001-8709-1088}
\email{zhangyiwen@ahu.edu.cn}
\affiliation{%
  \institution{Anhui University}
  \city{Hefei}
  \state{Anhui Province}
  \country{China}
}

\author{Yi Zhang}
\orcid{0000-0001-8196-0668}
\email{zhangyi.ahu@gmail.com}
\affiliation{%
  \institution{Anhui University}
  \city{Hefei}
  \state{Anhui Province}
  \country{China}
}

\author{Lei Sang}
\orcid{0009-0007-1480-6522}
\email{sanglei@ahu.edu.cn}
\affiliation{%
  \institution{Anhui University}
  \city{Hefei}
  \state{Anhui Province}
  \country{China}
}

\author{Jieming Zhu}
\orcid{0000-0002-5666-8320}
\email{jiemingzhu@ieee.org}
\affiliation{%
  \institution{Huawei Noah’s Ark Lab}
  \city{Shenzhen}
  \state{Guangdong Province}
  \country{China}
}

\renewcommand{\shortauthors}{Honghao Li, Yiwen Zhang, Yi Zhang, Lei Sang, \& Jieming Zhu}

\begin{abstract}
    Hadamard Product (HP) has long been a cornerstone in click-through rate (CTR) prediction tasks due to its simplicity, effectiveness, and ability to capture feature interactions without additional parameters. However, the underlying reasons for its effectiveness remain unclear. In this paper, we revisit HP from the perspective of Quadratic Neural Networks (QNN), which leverage quadratic interaction terms to model complex feature relationships. We further reveal QNN's ability to expand the feature space and provide smooth nonlinear approximations without relying on activation functions. Meanwhile, we find that traditional post-activation does not further improve the performance of the QNN. Instead, mid-activation is a more suitable alternative. Through theoretical analysis and empirical evaluation of 25 QNN neuron formats, we identify a good-performing variant and make further enhancements on it. Specifically, we propose the Multi-Head Khatri-Rao Product as a superior alternative to HP and a Self-Ensemble Loss with dynamic ensemble capability within the same network to enhance computational efficiency and performance. Ultimately, we propose a novel neuron format, QNN-$\alpha$, which is tailored for CTR prediction tasks. Experimental results show that QNN-$\alpha$ achieves new state-of-the-art performance on six public datasets while maintaining low inference latency, good scalability, and excellent compatibility. The code, running logs, and detailed hyperparameter configurations are available at: \url{https://github.com/salmon1802/QNN}.
\end{abstract}




\maketitle

\section{Introduction}
Hadamard Product (HP), known for its simplicity, effectiveness, and without additional parameter requirements, has long been one of the most popular explicit feature interaction methods for click-through rate (CTR) prediction tasks \cite{dcnv2, AFN, RFM, DCNv3}. By explicitly capturing bounded interactions between features, HP demonstrates strong computational efficiency and model performance in large-scale sparse data scenarios \cite{dcnv2}. Recent studies \cite{FINAL, RFM, DCNv3} have further achieved significant performance improvements by leveraging HP. However, despite the encouraging results, the underlying reasons for HP's effectiveness remain unclear, motivating us to further explore the fundamental causes of HP's success.

\begin{figure}[t]
    \centering
    \begin{minipage}[t]{0.75\linewidth}
        \centering
    \includegraphics[width=\linewidth]{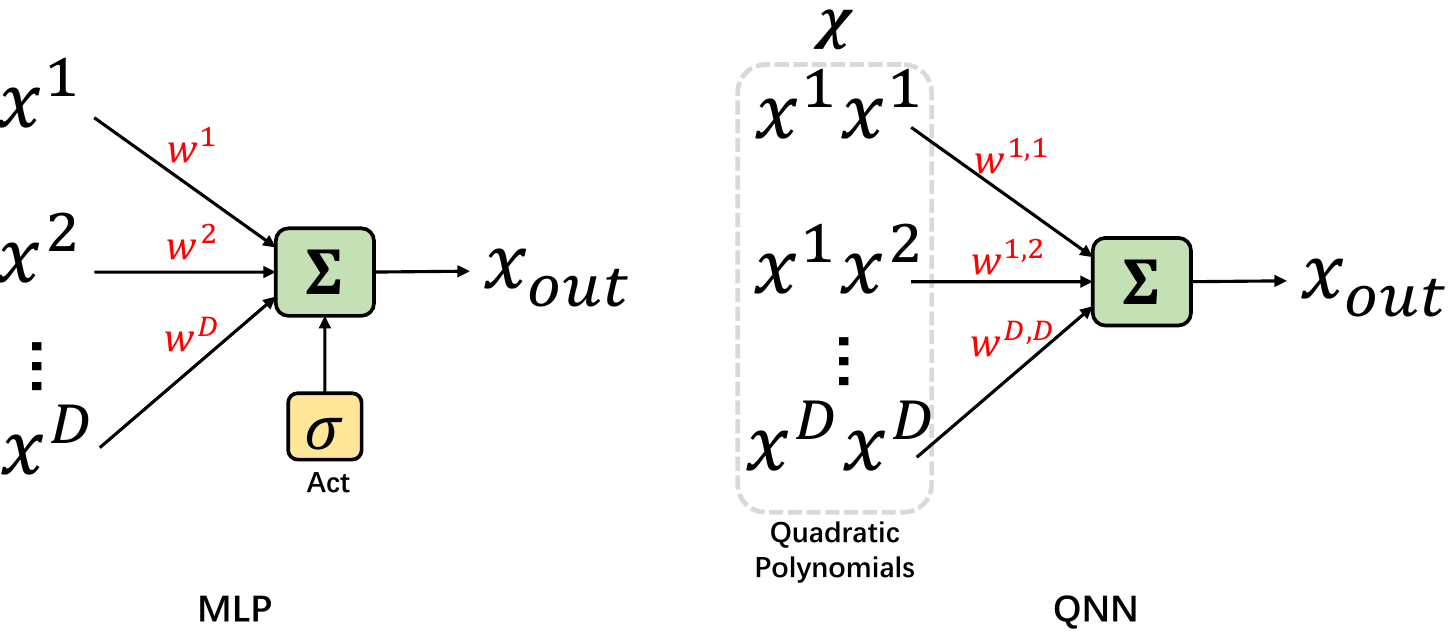}
    \end{minipage}
    \captionsetup{justification=raggedright, font=small}
    \caption{Comparison of MLP and QNN.}
    \label{introduction}
    \vspace{-2em}
\end{figure}

To investigate its nature, we attempt to analyze HP from the perspective of Quadratic Neural Networks (QNN) \cite{QNNbegin, qnn1, qnn2, Quadralib}. The element-wise multiplication operation of HP is naturally related to the core idea of QNN, which involves linearly independent quadratic polynomials $\chi$ (As shown in Figure \ref{introduction}). Therefore, we believe that revisiting the working mechanism of HP through the lens of QNN can provide new insights into CTR prediction. Specifically, we first derive and compare recursive formulations for MLP and HP-based CTR models from the theoretical foundation of QNN, aiming to deeply analyze the source of HP's superiority in capturing feature interactions. Through empirical evaluation and visualization analyses in Section \ref{2.1}, we observe that the nonlinear approximation ability of MLP is relatively sharp and \textit{heavily dependent on activation functions}. In contrast, the fundamental reason for HP's effectiveness lies in its expansion of the feature space and its inherent smooth nonlinear approximation capabilities, \textit{without explicitly relying on activation functions}. This discovery raises an intriguing question worthy of further exploration: \textit{Are Activation Functions Necessary for HP-based CTR Prediction?}

To answer this question, we empirically analyze 25 different neuron formats of QNN, as discussed in Section \ref{2.2} and \ref{2.3}, claim the connections between existing CTR models and QNN, and identify the T19 variant as a strong performer. Moreover, we find that, in most cases, introducing post-activation (Post Act) into QNN reduces performance, while introducing mid-activation (Mid Act) helps QNN achieve both smooth and sharp nonlinear approximation capabilities, thereby improving performance. Based on these findings, a follow-up question naturally arises: \textit{Is there a better QNN design for CTR prediction?}

In this paper, we deliver a positive answer to this question. Initially, we attempt to improve performance by increasing the parameter size of QNN neurons or modifying them into a bottleneck structure \cite{resNet}, but the results do not meet our expectations. Subsequently, we focus on the Product method for feature interactions and replace the traditional HP with a more expressive Khatri–Rao Product (KRP) \cite{Khatri-Rao}, achieving certain performance improvements. However, since KRP requires higher computational resources compared to HP, we propose the \textbf{Multi-Head Khatri–Rao Product (MH-KRP)}. Theoretically, MH-KRP reduces the parameter size and computational complexity of KRP by a factor of $H$.

Furthermore, while ensemble learning is an effective method to mitigate the risk of overfitting on sparse data in CTR prediction tasks, it often incurs high computational costs \cite{CTRensemble1, finalmlp, EKTF, PINTEREST}. To address this issue, we propose a \textbf{Self-Ensemble Loss (SE Loss)}, which dynamically ensembles the prediction results generated by the same network during different forward passes, thereby achieving performance improvements without additional inference costs. Ultimately, we propose a novel QNN format \textbf{QNN-$\alpha$} that is better suited for CTR prediction tasks. This model not only achieves state-of-the-art performance on six public datasets but also exhibits lower inference latency compared to some models that have already been successfully deployed in production environments. The core contributions of this paper are summarized as follows:
\begin{itemize}[leftmargin=*]
\item  To the best of our knowledge, this is the first work to introduce QNN into CTR prediction. From the perspective of QNN, we provide new insights into feature interaction-based CTR prediction.
\item We reveal the core reasons why HP improves the performance of CTR models through theoretical analysis and visualization. Moreover, we find that Post Act does not effectively enhance the performance of QNN, while Mid Act proves to be a better choice.
\item We evaluate 25 QNN neuron formats, propose MH-KRP as a superior alternative to HP, and introduce SE Loss to boost performance without extra sub-networks. Ultimately, we propose QNN-$\alpha$, a novel tailored format for CTR prediction.
\item We conduct extensive experiments on six benchmark datasets, demonstrating the low latency, effectiveness, scalability, and compatibility of QNN-$\alpha$.
\end{itemize}

\subsection{Preliminaries}
\subsubsection{\textbf{DEFINITION 1: CTR Prediction.}} It is a binary classification task that predicts the probability of a user clicking on an item using user profiles $x_U$, item attributes $x_I$, and context $x_C$ features \cite{openbenchmark, CETN}. A CTR input sample can be defined as a tuple: $X = \{x_U, x_I, x_C\}$, where $y \in \{0, 1\}$ represents the true label of user click behavior. The goal of a CTR prediction model is to predict $y$ and rank items based on the predicted probabilities $\hat{y}$. Most CTR models \cite{autoint, CETN, adagin} use an embedding layer to convert $X$ into low-dimensional dense vectors: $\mathbf{e}_i = \textit{E}_i x_i$, where $\textit{E}_i \in \mathbb{R}^{d \times s_i}$ is the embedding matrix, $s_i$ is the vocabulary size for the $i$-th field, and $d$ is the embedding dimension. After that, we concatenate the individual feature embeddings to get the input $\mathbf{X}_1=\left[\mathbf{e}_1, \mathbf{e}_2, \cdots, \mathbf{e}_f\right] \in \mathbb{R}^D$, where $D=\sum_{i=1}^f d$, and $f$ denotes the number of feature fields. In this work, we take the input $\mathbf{X}_1$ as the first-order feature.

\subsubsection{\textbf{DEFINITION 2: Feature Interaction.}} 
\textit{Implicit feature interaction} leverages MLP to automatically learn high-order feature interactions \cite{xdeepfm, dcnv2}, while \textit{explicit feature interaction} directly models relationships between features via explicit formulas \cite{xdeepfm, dcnv2}. A common explicit method is $\mathbf{X}_n = \mathbf{X}_1 \odot \mathbf{X}_{n-1}$ \cite{dcnv2, DCNv3}, which uses the Hadamard Product to generate the $n$-th order feature $\mathbf{X}_n$.

\subsubsection{\textbf{DEFINITION 3: Quadratic Neural Networks}}
The core idea of QNN \cite{QNNbegin} is that the output of neurons in each layer depends not only on the linear transformation of input features, but also explicitly introduces quadratic interaction terms between input features \cite{qnn1,Quadralib,infinite}. Specifically, QNN constructs the feature space using a linearly independent set of quadratic polynomials $\chi = \{x^1x^1, x^1x^2, \dots\}$, where $x^i$ indicate $i$-th variable, $\chi$ includes all possible quadratic interaction terms. This approach enhances feature relationship explicit modeling and shortens the learning path for complex interactions \cite{DCNv3}.

\section{Investigation of Feature Interactions in CTR}
\label{revis}
\subsection{Revisiting Feature Interactions in CTR from Quadratic Neural Networks}
\label{2.1}
\begin{figure*}[t]

    \centering
    \begin{minipage}[t]{0.16\linewidth}
        \centering
        \includegraphics[width=\linewidth]{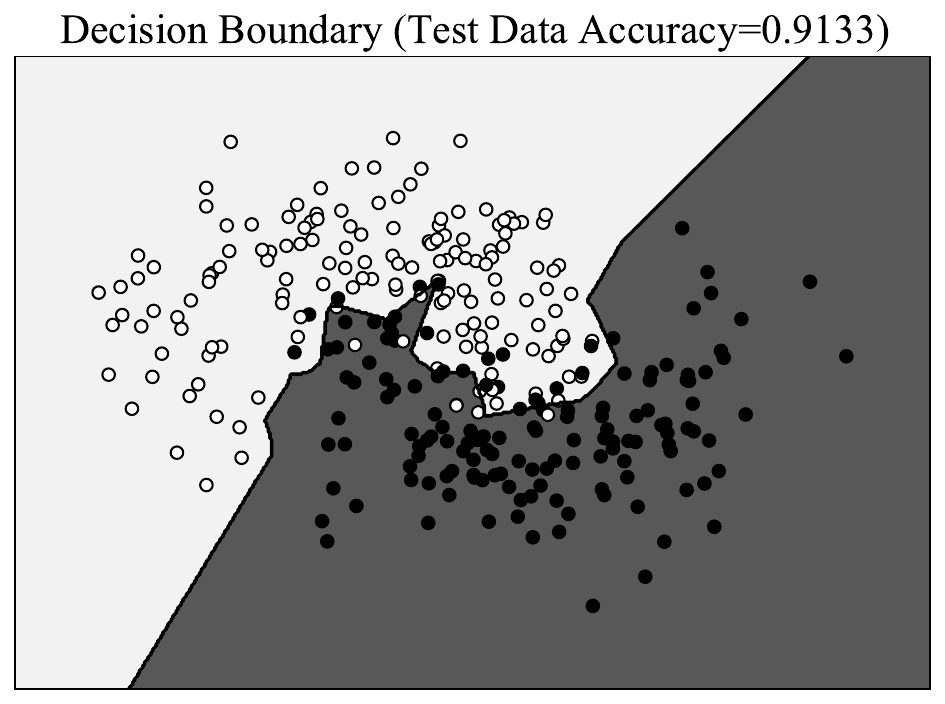}
    \end{minipage}
    \begin{minipage}[t]{0.16\linewidth}
        \centering
        \includegraphics[width=\linewidth]{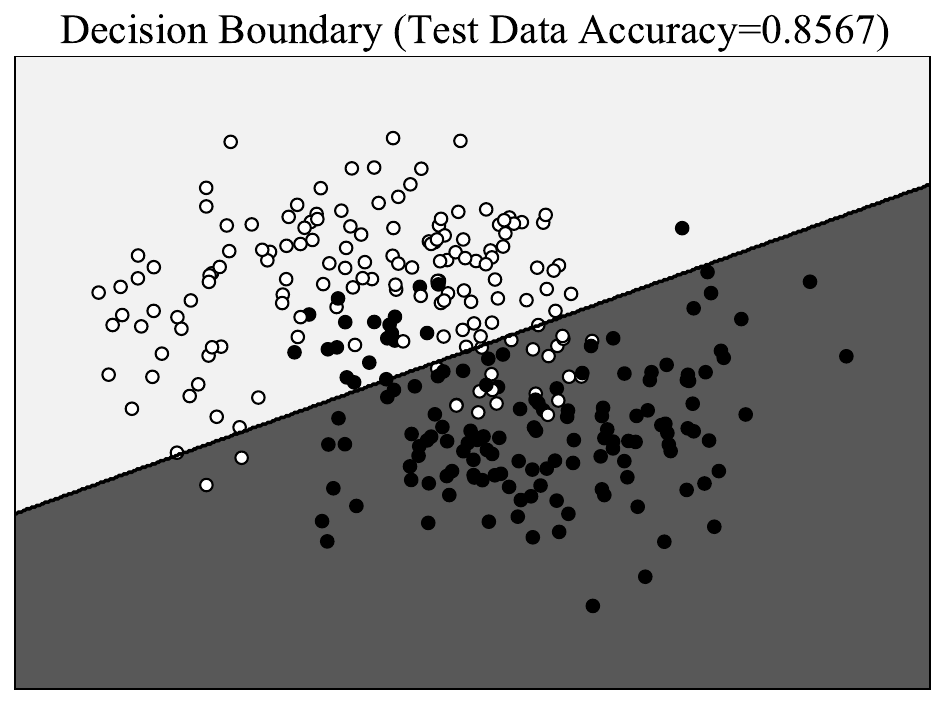}
    \end{minipage}
    \begin{minipage}[t]{0.16\linewidth}
        \centering
        \includegraphics[width=\linewidth]{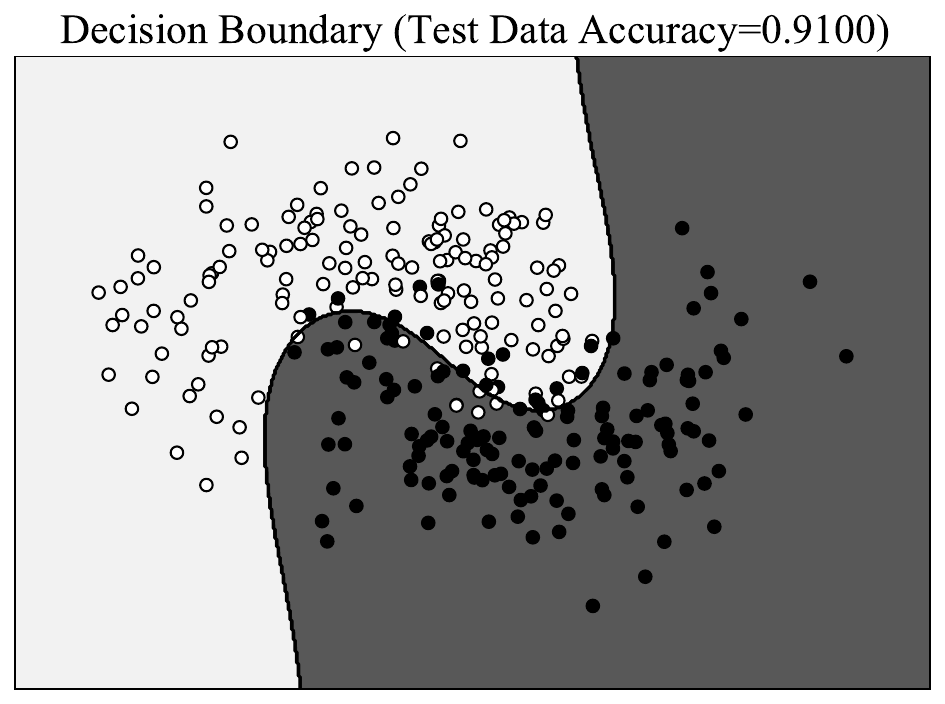}
    \end{minipage}
    \begin{minipage}[t]{0.16\linewidth}
    \includegraphics[width=\linewidth]{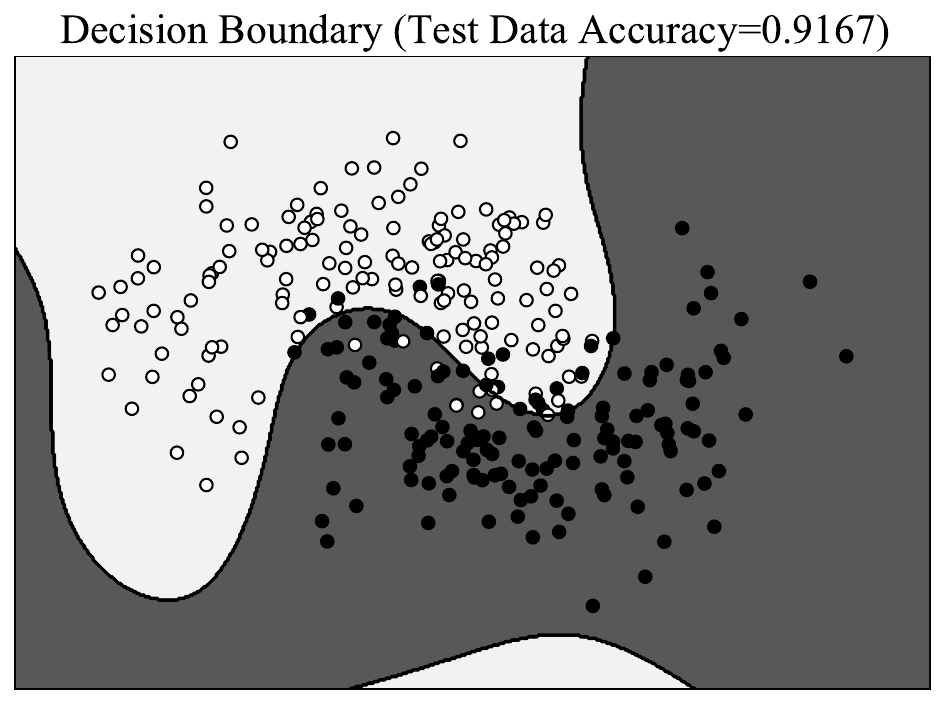}
    \end{minipage}
    \begin{minipage}[t]{0.16\linewidth}
    \includegraphics[width=\linewidth]{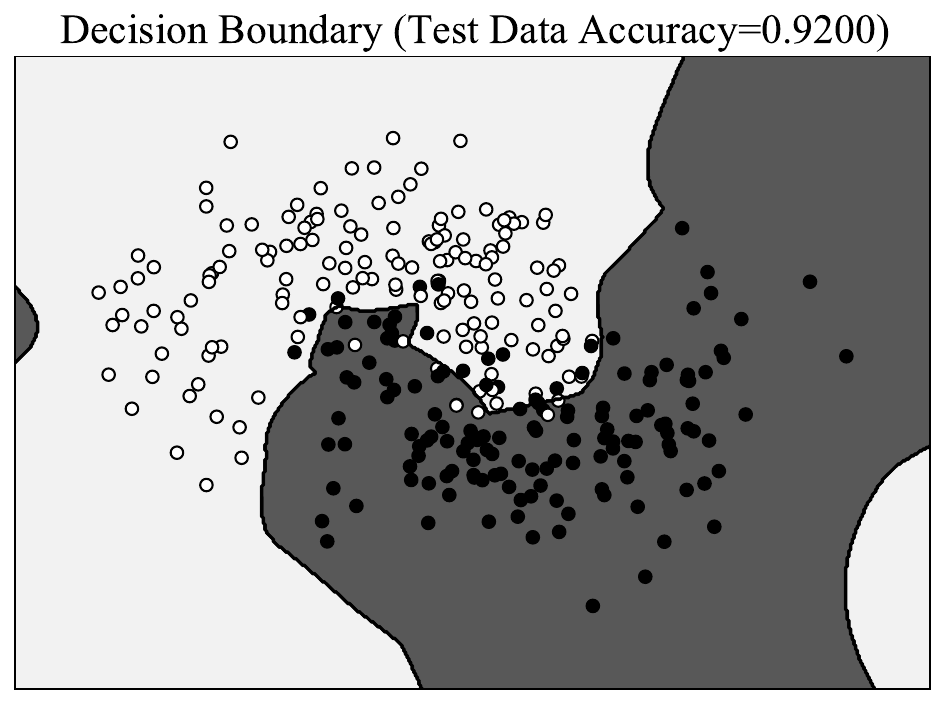}
    \end{minipage} %
    \begin{minipage}[t]{0.16\linewidth}
    \includegraphics[width=\linewidth]{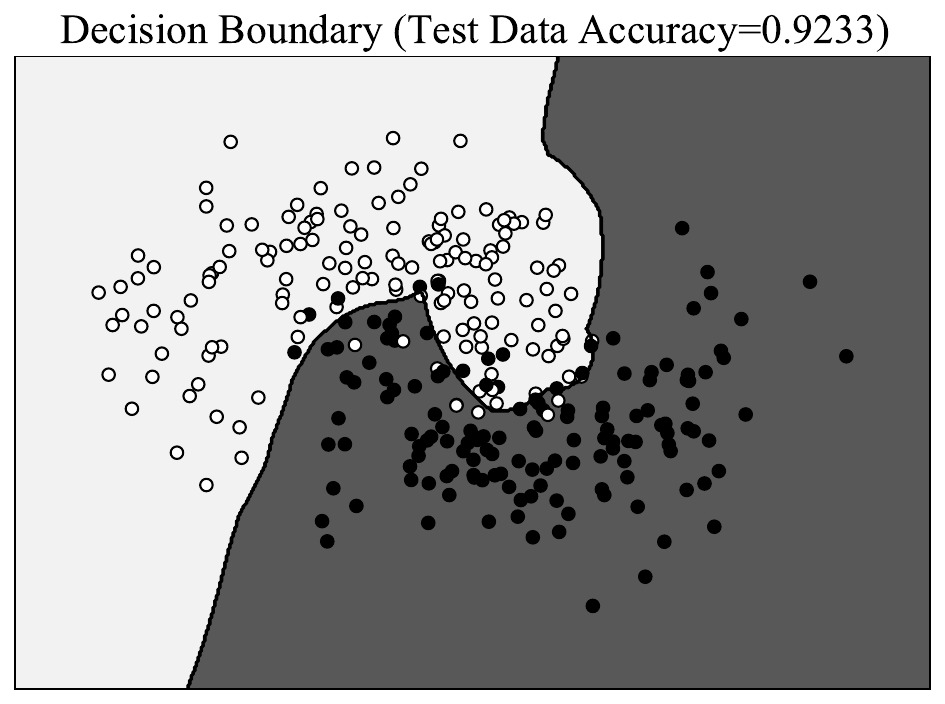}
    \end{minipage} 

    \begin{minipage}[t]{0.16\linewidth}
        \centering
        \includegraphics[width=\linewidth]{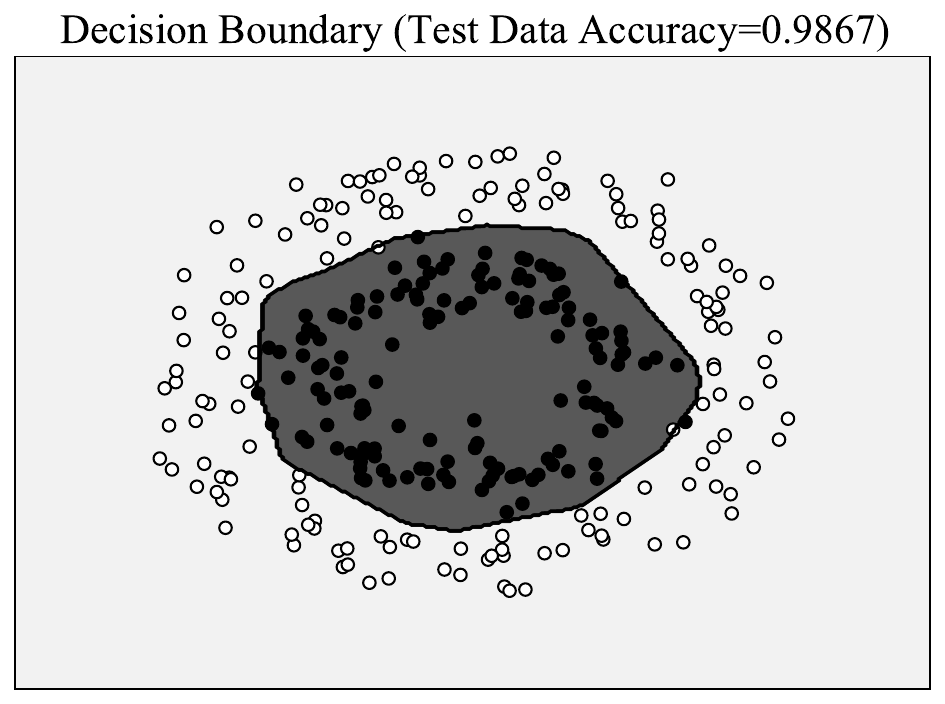}
        \subcaptionsetup{font=footnotesize}
        \subcaption{MLP}
    \end{minipage}
    \begin{minipage}[t]{0.16\linewidth}
        \centering
        \includegraphics[width=\linewidth]{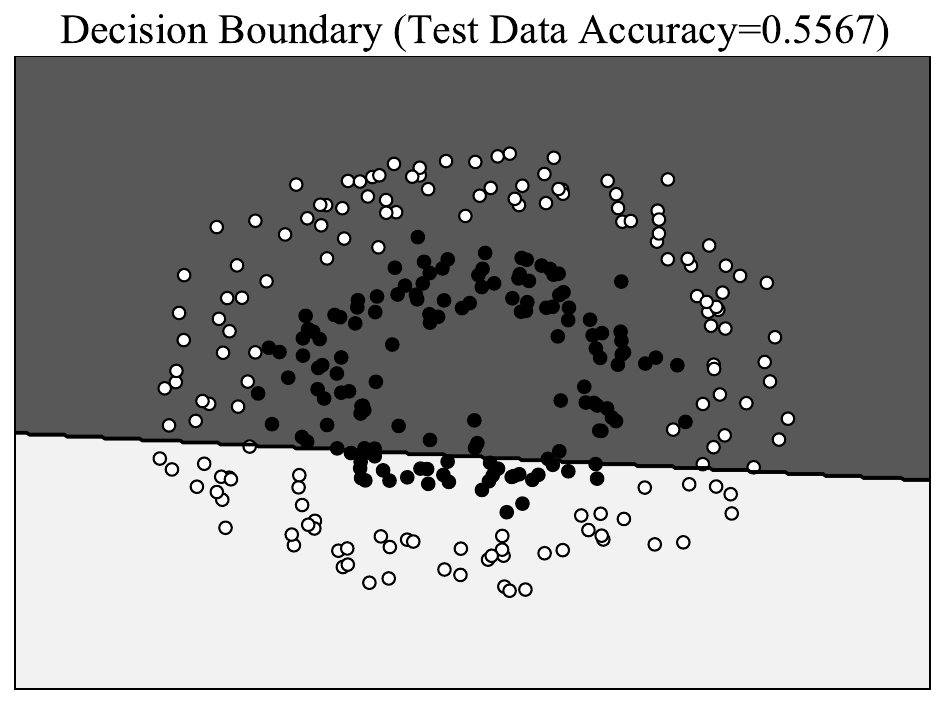}
        \subcaptionsetup{font=footnotesize}
        \subcaption{MLP w/o Act}
    \end{minipage}
    \begin{minipage}[t]{0.16\linewidth}
        \centering
        \includegraphics[width=\linewidth]{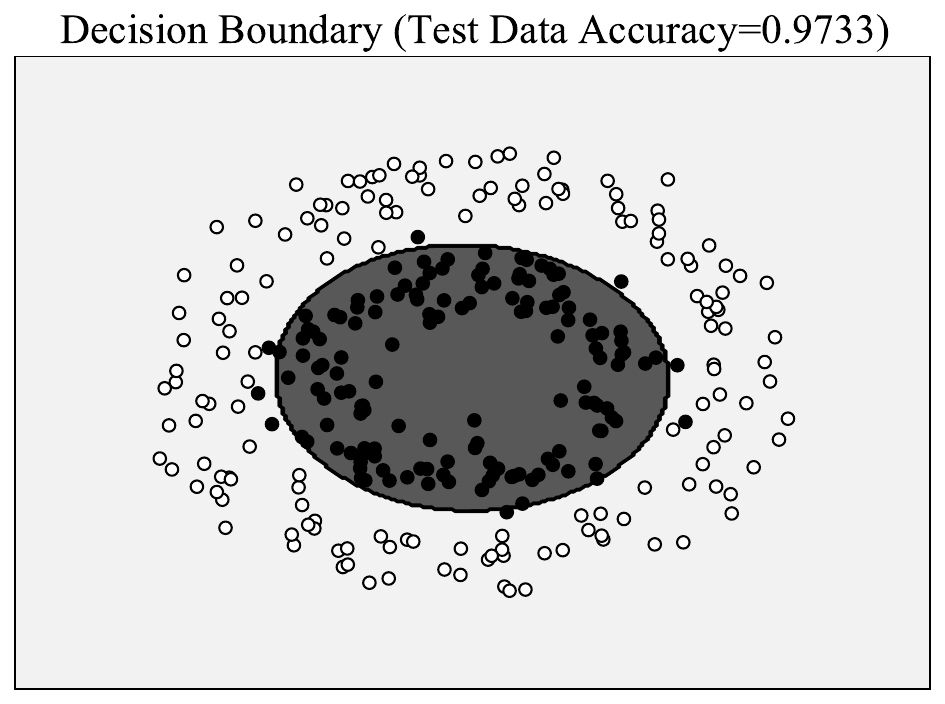}
        \subcaptionsetup{font=footnotesize}
        \subcaption{CrossNetv2 (CNv2)}
    \end{minipage}
    \begin{minipage}[t]{0.16\linewidth}
    \includegraphics[width=\linewidth]{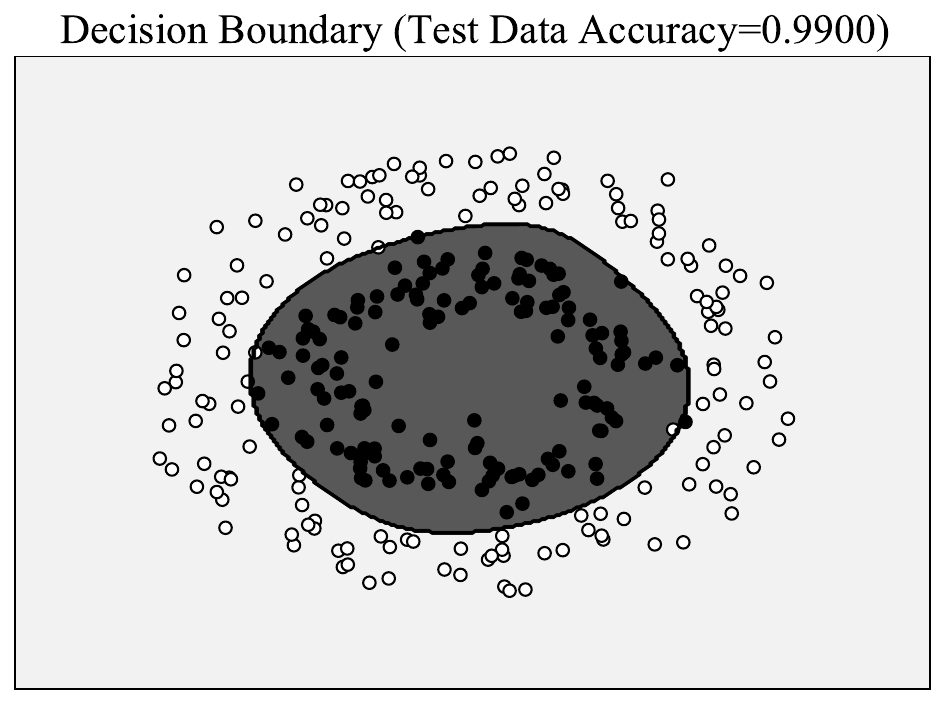}
    \subcaptionsetup{font=footnotesize}
    \subcaption{T9 (ECN-like)}
    \end{minipage}
    \begin{minipage}[t]{0.16\linewidth}
    \includegraphics[width=\linewidth]{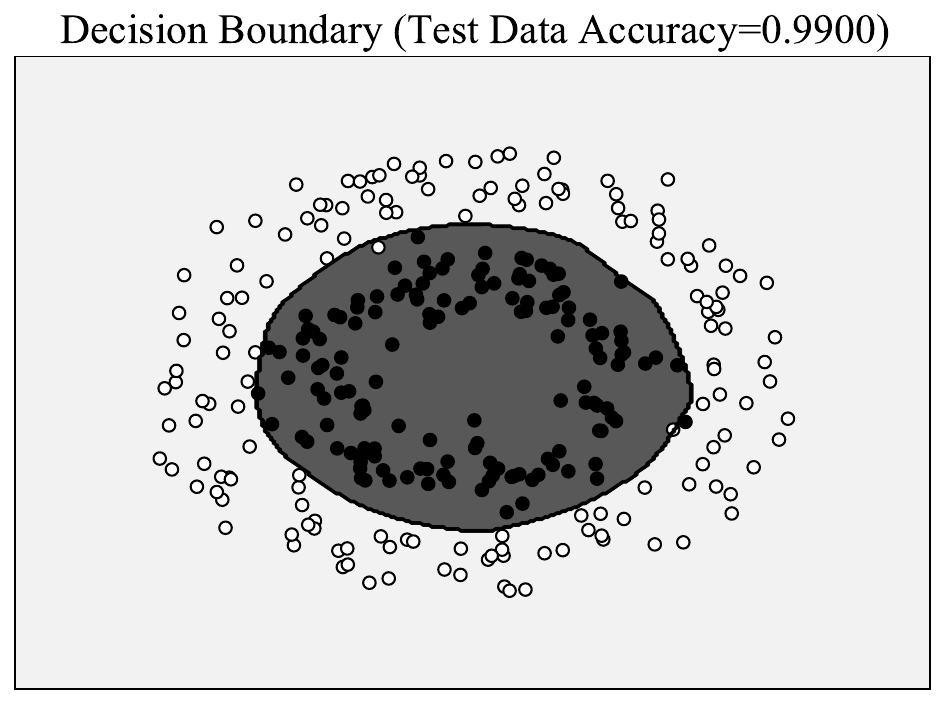}
    \subcaptionsetup{font=footnotesize}
    \subcaption{DCNv2=MLP+CNv2}
    \end{minipage} %
    \begin{minipage}[t]{0.16\linewidth}
    \includegraphics[width=\linewidth]{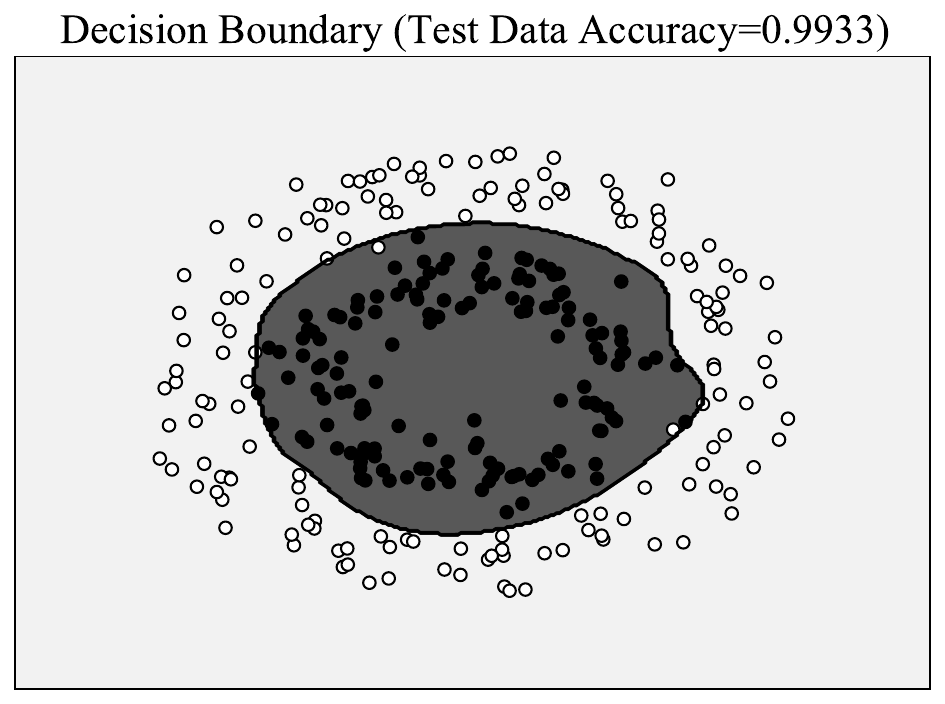}
    \subcaptionsetup{font=footnotesize}
    \subcaption{T19}
    \end{minipage} 
    \captionsetup{justification=raggedright, font=small}
    \caption{Comparison of decision boundaries for 2D moon (line 1) and circle (line 2) datasets in different neuron formats.}
    \label{db}
    \vspace{-1em}
\end{figure*}

\subsubsection{\textbf{Theoretical analysis}} CTR prediction models typically enhance performance by integrating explicit and implicit feature interactions \cite{openbenchmark, xdeepfm}. In this paper, we revisit several classic methods for feature interactions \cite{dcnv2, DNN}, which assume that explicit and implicit feature interactions are distinct yet complementary \cite{xdeepfm, dcnv2}. Firstly, we investigate the classic DCNv2 \cite{dcnv2} model, which employs CrossNetv2 (CNv2) for explicit feature interactions. Formally, the recursive formula in CNv2 is defined as follows\footnote{ The bias and residual terms are omitted for clarity.}:
\begin{equation}
\label{CrossNetv2}
\begin{aligned}
\mathbf{X}_{l+1} &= \mathbf{X}_1 \odot \mathbf{W}_l \mathbf{X}_l,\ \ l=1,2,\dots,L, \\
\end{aligned}
\end{equation}
where $\mathbf{W}_{l} \in \mathbb{R}^{D \times D}$ is the learnable parameter matrix, $\mathbf{X}_{l} \in \mathbb{R}^{D}$ represents the $l$-th order feature, $L$ is the total number of layers, and $\odot$ denotes the Hadamard Product. In fact, we can rewrite the Equation (\ref{CrossNetv2}) as:
\begin{equation}
\label{Ditem}
\begin{aligned}
x_{l+1}^i = x_1^i \sum_{j=1}^{D} w_l^{i,j} x_l^j = \underbrace{w_l^{i,1} x_1^i x_l^1 + w_l^{i,2} x_1^i x_l^2 + \cdots + w_l^{i,D} x_1^i x_l^D}_{D \text { interaction items }},
\end{aligned}
\end{equation}
where $w_{l}^{i,j} \in \mathbf{W}_l$ represents the learnable parameter at position $(i, j)$, and $x_{l}^j \in \mathbf{X}_l$ denotes the $j$-th element of $\mathbf{X}_l$. From Equation (\ref{Ditem}), we observe that the recursive function proposed by CNv2 expands the 1-dimensional feature space into a $D$-dimensional feature space, where second-order polynomials $\chi=\{x_1^i x_l^1, x_1^i x_l^2, \cdots, x_1^i x_l^D\}$ introduces a nonlinear dependency between the variables $x_1^i$ and $x_l^j$.  Therefore, CNv2 can be regarded as a QNN. Similarly, the MLP for implicit feature interactions can be rewritten as:
\begin{equation}
\label{mlp}
\begin{aligned}
x_{l+1}^i = \sigma\left(\sum_{j=1}^{D} w_l^{i,j} x_l^j\right)= \underbrace{\sigma\left(w_l^{i,1} x_l^1 + w_l^{i,2} x_l^2 + \cdots + w_l^{i,D} x_l^D\right)}_{D \text { interaction items }},
\end{aligned}
\end{equation}
where $\sigma$ represents the activation function (Act). It is evident that, without considering the activation function, although Equation (\ref{mlp}) also contains $D$ interaction terms, it does not include the product of multiple variables, resulting in a linear correlation with $x_l^j$. Therefore, the non-linear approximation ability of the MLP heavily depends on the Act \cite{MLPapproximators}.

ECN \cite{DCNv3} adopts a similar method to CNv2 for modeling feature interactions. The difference lies in that the former is designed to capture exponentially growing feature interactions, while the latter can only capture linearly growing feature interactions. The recursive function of ECN-like layers, which we refer to as the T9 neuron format in this paper, can be simplified as follows:
\begin{equation}
\label{ECN}
\begin{aligned}
\mathbf{X}_{l+1} &= \mathbf{X}_l \odot \mathbf{W}_l \mathbf{X}_l \overset{l\ >\ 1}{=} \mathbf{X}_{l-1} \odot \mathbf{W}_{l-1} \mathbf{X}_{l-1} \odot \mathbf{W}_l \mathbf{X}_l, \\
x_{l+1}^i &\overset{l\ >\ 1}{=} x_{l-1}^i \sum_{j=1}^{D} w_{l-1}^{i,j} x_{l-1}^j \sum_{k=1}^{D} w_l^{i,k} x_l^k, \\
&= \underbrace{A_{(1,1)} x_{l-1}^1 x_l^1 + A_{(1,2)} x_{l-1}^1 x_l^2 + \cdots + A_{(D,D)} x_{l-1}^D x_l^D}_{D^2 \text { interaction items}},
\end{aligned}
\end{equation}
where $A$ is a simplification coefficient, defined as:
\begin{equation}
\label{A}
\begin{aligned}
A_{(j, k)} = w_{l-1}^{i,j} w_{l}^{i,k} x_{l-1}^i,
\end{aligned}
\end{equation}
From the perspective of QNN, ECN's second-order polynomials are defined as $\chi=\{x_{l-1}^1 x_l^1, x_{l-1}^1 x_l^2, \cdots, x_{l-1}^D x_l^D\}$. We observe that ECN expands the feature space from $D$ dimensions in CNv2 to $D^2$ dimensions. Moreover, each interaction term exhibits a nonlinear relationship with $x_{l-1}^i$. As a result, ECN may have better nonlinear approximation capabilities than CNv2.

\subsubsection{\textbf{Visual Analysis of Nonlinear Approximation Ability}}
To more intuitively analyze and compare the approximation capabilities between QNN and MLP, we visualize the decision boundaries of various neuron formats on the widely-used 2D moon and circle test dataset\footnote{ For hyperparameter configurations, we uniformly set the hidden dimension size to 20, the learning rate to 1e-2, the number of layers to 2, and training 500 epochs on moon dataset and 100 epochs on circle dataset.} \cite{scikit} as shown in Figure \ref{db}. We observe that, except for Figure \ref{db} (b), all other neuron formats exhibit nonlinear approximation capabilities, which aligns with our expected results and validates the correctness of our theoretical analysis.

Furthermore, MLP shows sharper decision boundaries compared to QNN variants, CNv2 and T9. This sharpness may be due to the fact that the ReLU function is not smooth at zero \cite{swish, mish}. Meanwhile, CNv2 and T9 exhibit extremely smooth nonlinear decision boundaries, with T9 demonstrating accuracy surpassing that of MLP on two datasets. Thus, we conclude that relying solely on the nonlinear approximation capability of ReLU often results in sharper decision boundaries. In contrast, QNN inherently possesses a smooth nonlinear approximation capability, which may potentially help reduce misclassifications near the boundaries.

Besides, we observe that DCNv2, which combines both MLP and CNv2, blends the sharp and smooth decision boundaries, achieving higher accuracy. This performance enhancement has been validated by previous works \cite{dcnv2, xdeepfm}, where explicit (e.g., CNv2) and implicit (e.g., MLP) feature interactions are combined to leverage the complementary strengths of different interaction methods. In Figure \ref{db}, this abstract concept of explicit and implicit can be further intuitively understood: there is often a need for sharp boundaries with fine localized carving, as well as boundaries that allow for globally smooth transitions.

\begin{table*}[t]
\renewcommand\arraystretch{1}
\captionsetup{font=footnotesize}
\caption{Performance comparison of QNN on Criteo dataset. CTR researchers consider a \textit{0.1\%} improvement in Logloss and AUC as statistically significant \cite{openbenchmark}.}
\label{NF}
\resizebox{1.0\linewidth}{!}{
\begin{tabular}{c|c|cc|c|cc|c|c|cc}
\hline
\multirow{2}{*}{\textbf{Type}} &
  \multirow{2}{*}{\textbf{Neuron Format}} &
  \multicolumn{2}{c|}{\textbf{Post Act:\ ReLU($\Phi(\mathbf{X})$)}} &
  \multirow{2}{*}{\textbf{Time×Epochs}} &
  \multicolumn{2}{c|}{\textbf{W/O Post Act:\ $\Phi(\mathbf{X})$}} &
  \multirow{2}{*}{\textbf{Time×Epochs}} &
  \multirow{2}{*}{\textbf{\#Params}} &
  \multicolumn{2}{c}{\textbf{Gap}} \\ \cline{3-4} \cline{6-7} \cline{10-11} 
 &
   &
  Logloss$\downarrow$ &
  AUC(\%)$\uparrow$ &
   &
  Logloss$\downarrow$ &
  AUC(\%)$\uparrow$ &
   &
   &
 $\triangle$Logloss&
 $\triangle$AUC \\ \hline
\rowcolor{red!7} MLP \cite{DNN}&
  $\Phi(\mathbf{X}) = \mathbf{W}_\textit{a} \mathbf{X}$ &
  0.4381 &
  81.37 &
  2min × 11 &
  0.4568 &
  79.32 &
  2min × 12 &
  15.14M &
  +0.0187 &
  -2.05 \\
\rowcolor{cyan!7} CrossNetv2 \cite{dcnv2}&
  $\Phi(\mathbf{X}) = \mathbf{X}_1 \odot \mathbf{W}_\textit{a} \mathbf{X} + \mathbf{X}$ &
  0.4389 &
  81.34 &
  2min × 12 &
  0.4385 &
  81.36 &
  2min × 10 &
  15.74M &
  -0.0004 &
  +0.02 \\  \hline
\rowcolor{cyan!7} T1 (FM-like \cite{FM, finalmlp}) &
  $\Phi(\mathbf{X}) = \mathbf{X}^\intercal \mathbf{W}_\textit{a} \mathbf{X} + \mathbf{W}_\textit{b} \mathbf{X}$ &
  0.4395 &
  81.27 &
  24min × 11 &
  0.4388 &
  81.36 &
  24min × 10 &
   17.66M &
  -0.0007 &
  +0.09 \\
\rowcolor{red!7} T2 &
  $\Phi(\mathbf{X}) = \mathbf{X}^\intercal \mathbf{W}_\textit{a} \mathbf{X}$ &
  0.4391 &
  81.32 &
  23min × 10 &
  0.4439 &
  81.30 &
  23min × 14 &
  17.63M &
  +0.0048 &
  -0.02 \\
\rowcolor{cyan!7} T3 &
  $\Phi(\mathbf{X}) = \mathbf{W}_\textit{a} \mathbf{X}^\text{2}$ &
  0.4401 &
  81.22 &
  2min × 16 &
  0.4412 &
  81.26 &
  2min × 31 &
  15.74M &
  +0.0011 &
  +0.04 \\
\rowcolor{cyan!7} T4 &
  $\Phi(\mathbf{X}) = (\mathbf{W}_\textit{a} \mathbf{X})^\text{2}$ &
  0.4382 &
  81.39 &
  2min × 16 &
  0.4381 &
  81.41 &
  2min × 12 &
  15.74M &
  -0.0001 &
  +0.02 \\
\rowcolor{red!7} T5 (PEPNet-like \cite{PEPNET})&
  $\Phi(\mathbf{X}) = \mathbf{W}_\textit{a} \mathbf{X} \odot \mathbf{W}_\textit{b} \mathbf{X}$ &
  0.4383 &
  81.39 &
  2min × 11 &
  0.4385 &
  81.37 &
  2min × 9 &
  16.91M &
  +0.0002 &
  -0.02 \\
\rowcolor{red!7} T6 &
  $\Phi(\mathbf{X}) = \mathbf{X}^\intercal \mathbf{W}_\textit{a} \mathbf{X} + \mathbf{W}_\textit{b} \mathbf{X}^\text{2}$ &
  0.4416 &
  81.28 &
  23min × 13 &
  0.4468 &
  81.22 &
  23min × 15 &
  17.66M &
  +0.0052 &
  -0.06 \\
\rowcolor{red!7} T7 &
  $\Phi(\mathbf{X}) = \mathbf{W}_\textit{a} \mathbf{X} \odot \mathbf{W}_\textit{b} \mathbf{X} + \mathbf{W}_\textit{c} \mathbf{X}^\text{2}$ &
  0.4391 &
  81.28 &
  2.5min × 11 &
  0.4403 &
  81.24 &
  2.5min × 11 &
  18.08M &
  +0.0012 &
  -0.04 \\
\rowcolor{green!7} T8 &
  $\Phi(\mathbf{X}) = \mathbf{W}_\textit{a} \mathbf{X} \odot \mathbf{W}_\textit{b} \mathbf{X} + \mathbf{W}_\textit{c} \mathbf{X}$ &
  0.4384 &
  81.37 &
  2.5min × 11 &
  0.4384 &
  81.37 &
  2.5min × 13 &
  18.08M &
  0.0000 &
  0.00 \\
\rowcolor{cyan!7} T9 (ECN-like \cite{DCNv3})&
  $\Phi(\mathbf{X}) = \mathbf{X} \odot \mathbf{W}_\textit{a} \mathbf{X} + \mathbf{X}$  &
  \textbf{0.4369} &
  \textbf{81.51} &
  2.5min × 17 &
  \textbf{0.4369} &
  \textbf{81.53} &
  2.5min × 16 &
  15.74M &
  0.0000 &
  +0.02 \\
\rowcolor{cyan!7} T10 &
  $\Phi(\mathbf{X}) = \mathbf{X} \odot \mathbf{W}_\textit{a} \mathbf{X} + \mathbf{W}_\textit{b} \mathbf{X}$  &
  0.4381 &
  81.39 &
  2.5min × 10 &
  0.4373 &
  81.50 &
  2.5min × 19 &
  16.91M &
  -0.0008 &
  +0.11 \\
\rowcolor{cyan!7} T11 &
  $\Phi(\mathbf{X}) = \mathbf{W}_\textit{a} \mathbf{X}\ ||\ (\mathbf{W}_\textit{b} \mathbf{X})^2$  &
  0.4381 &
  81.38 &
  2min × 12 &
  0.4378 &
  81.44 &
  2min × 12 &
  15.74M &
  -0.0003 &
  +0.06 \\
\rowcolor{red!7} T12 &
  $\Phi(\mathbf{X}) = \mathbf{W}_\textit{a} \mathbf{X} \odot \mathbf{W}_\textit{b} \mathbf{X}\ ||\ \mathbf{W}_\textit{c} \mathbf{X}^2$ &
  0.4381 &
  81.40 &
  2.5min × 12 &
  0.4384 &
  81.35 &
  2.5min × 12 &
  16.32M &
  +0.0003 &
  -0.05 \\
\rowcolor{cyan!7} T13 (FINAL-like \cite{FINAL})&
  $\Phi(\mathbf{X}) = \mathbf{W}_\textit{a} \mathbf{X} \odot \mathbf{W}_\textit{b} \mathbf{X}\ ||\ \mathbf{W}_\textit{c} \mathbf{X}$ &
  0.4386 &
  81.35 &
  2min × 11 &
  0.4382 &
  81.39 &
  2min × 12 &
  16.32M &
  -0.0004 &
  +0.04 \\
\rowcolor{cyan!7} T14 &
  $\Phi(\mathbf{X}) = \mathbf{W}_\textit{a} \mathbf{X} \odot \mathbf{W}_\textit{b} \mathbf{X}\ ||\ \mathbf{W}_\textit{b} \mathbf{X}$ &
  0.4387 &
  81.34 &
  2.5min × 15 &
  0.4379 &
  81.42 &
  2.5min × 12 &
  15.74M &
  -0.0008 &
  +0.08 \\
\rowcolor{cyan!7} T15 &
  $\Phi(\mathbf{X}) = \mathbf{W}_\textit{a} \mathbf{X} \ ||\ (\mathbf{W}_\textit{a} \mathbf{X})^2$ &
  0.4377 &
  81.42 &
  2min × 15 &
  0.4378 &
  81.44 &
  2min × 13 &
  15.15M &
  +0.0001 &
  +0.02 \\
\rowcolor{cyan!7} T16 (GCN-like \cite{GDCN})&
  $\Phi(\mathbf{X}) = \mathbf{X} \odot (\mathbf{W}_\textit{a} \mathbf{X} \odot \mathbf{W}_\textit{b} \mathbf{X}) + \mathbf{X}$ &
  0.4379 &
  81.43 &
  2.5min × 17 &
  0.4378 &
  81.48 &
  2.5min × 14 &
  16.91M &
  -0.0001 &
  +0.05 \\
\rowcolor{cyan!7} T17 &
  $\Phi(\mathbf{X}) = \mathbf{X} \odot (\mathbf{W}_\textit{a} \mathbf{X} + \mathbf{W}_\textit{b} \mathbf{X}) + \mathbf{X}$ &
  0.4384 &
  81.35 &
  2.5min × 13 &
  0.4372 &
  81.50 &
  2.5min × 15 &
  16.91M &
  -0.0012 &
  +0.15 \\
\rowcolor{cyan!7} T18 &
  $\Phi(\mathbf{X}) = \mathbf{X} \odot \mathbf{W}_\textit{a} \mathbf{X}\ ||\ \mathbf{X}$  &
  0.4392 &
  81.27 &
  2.5min × 10 &
  0.4388 &
  81.33 &
  2.5min × 10 &
  18.08M &
  -0.0004 &
  +0.06 \\
\rowcolor{cyan!7} T19 (as Figure \ref{db} (f)) &
  $\Phi(\mathbf{X}) = \mathbf{X} \odot \textbf{ReLU}(\mathbf{W}_\textit{a} \mathbf{X}) + \mathbf{X}$ &
  0.4377 &
  81.46 &
  2.5min × 18 &
  \textbf{0.4365} &
  \textbf{81.56} &
  2.5min × 26 &
  15.74M &
  -0.0012 &
  +0.10 \\
\rowcolor{red!7} T20 &
  $\Phi(\mathbf{X}) = \mathbf{X} \odot \mathbf{W}_\textit{a} \mathbf{X} + \mathbf{W}_\textit{a} \mathbf{X} + \mathbf{X}$  &
  0.4385 &
  81.34 &
  2.5min × 14 &
  0.4394 &
  81.26 &
  2.5min × 13 &
  15.74M &
  +0.0009 &
  -0.08 \\
\rowcolor{cyan!7} T21 &
  $\Phi(\mathbf{X}) = (\mathbf{X} \odot \mathbf{W}_\textit{a} \mathbf{X})^2 + \mathbf{X}$ &
  0.4430 &
  81.37 &
  2.5min × 30 &
  0.4382 &
  81.40 &
  2.5min × 22 &
  15.74M &
  -0.0048 &
  +0.03 \\
\rowcolor{cyan!7} T22 &
  $\Phi(\mathbf{X}) = \mathbf{X} \odot \mathbf{W}_\textit{a} \mathbf{X} + \alpha \odot \mathbf{X}$ &
  \textbf{0.4369} &
  \textbf{81.51} &
  2.5min × 19 &
  \textbf{0.4369} &
  \textbf{81.53} &
  2.5min × 16 &
  15.74M &
  0.0000 &
  +0.02 \\
\rowcolor{red!8} T23 &
  $\Phi(\mathbf{X}) = \mathbf{W}_\textit{a} \mathbf{X} \odot \mathbf{W}_\textit{b} \mathbf{X} + \mathbf{X}$ &
  0.4379 &
  81.42 &
  2.5min × 10 &
  0.4384 &
  81.41 &
  2.5min × 10 &
  16.91M &
  +0.0005 &
  -0.01 \\
\rowcolor{green!7} T24 (MaskNet-like \cite{masknet}) &
  $\Phi(\mathbf{X}) = \mathbf{W}_\textit{a} \mathbf{X} \odot \textbf{ReLU}(\mathbf{W}_\textit{b} \mathbf{X}) + \mathbf{X}$ &
  0.4378 &
  81.42 &
  2.5min × 12 &
  0.4381 &
  81.42 &
  2.5min × 12 &
  16.91M &
  +0.0003 &
  0.00 \\
\rowcolor{cyan!7} T25 &
  $\Phi(\mathbf{X}) = \mathbf{X} \odot \mathbf{W}_\textit{a}(\textbf{ReLU}(\mathbf{W}_\textit{b} \mathbf{X})) + \mathbf{X}$ &
  0.4378 &
  81.43 &
  2.5min × 14 &
  0.4378 &
  81.47 &
  2.5min × 14 &
  15.74M &
  0.0000 &
  +0.04 \\ \hline
\end{tabular}}
\vspace{-1em}
\end{table*}

\subsection{Necessity of Activation Functions}
\label{2.2}
The above analysis demonstrates that QNN possesses nonlinear approximation capabilities even without the use of activations. The question then arises: \textit{does the introduction of Activations further enhance the network's approximation ability?} To answer this question, we conduct experiments\footnote{ All networks share the same hyperparameters, with 3 layers fixed. $\triangle$AUC = AUC in $\Phi(\mathbf{X})$ - AUC in \textbf{ReLU}($\Phi(\mathbf{X})$), and $\triangle$Logloss is defined similarly. The \textbf{Gap} column compares performance with and without Post Act. To prevent overfitting,
we employ early stopping with a patience value of 2. Further details are in our open-source code and logs.} on the well-known large-scale Criteo \cite{openbenchmark} dataset. The experimental results are shown in Table \ref{NF}, where we refer to the classic QNN formats from \cite{Quadralib} and further extend them to 25 types. Meanwhile, we find that some formats correspond to existing CTR model designs, such as T1, T5, T9, T13, and T16.

Surprisingly, the $\Phi(\mathbf{X})$ variant outperforms \textbf{ReLU}($\Phi(\mathbf{X})$) in 64\% of cases (in blue) in terms of AUC, shows comparable performance in 8\% of cases (in green), and the former performs worse than the latter in 28\% of cases (in red). Even in the few cases where performance degradation occurs (e.g., T2, T6, and T7), the impact is minimal and significantly smaller than the performance loss observed when Post Act is removed in MLP. This indicates that QNN's inherent approximation ability is sufficient, making the use of Post Act unnecessary since it brings none or negligible benefits. In fact, Andrew \textit{et al.} \cite{LeakyReLU} have already pointed out that Act may lead to a certain degree of information loss due to its zero output for negative inputs, further highlighting the potential of QNN in CTR prediction tasks. Additionally, we observe that the training costs of networks with and without the Post Act have their advantages and disadvantages. For instance, in T4, the latter outperforms the former, while in T3, the opposite is true. Therefore, we conclude that the \textbf{Post Act} is unnecessary for QNN.

\subsection{Finding An Effective QNN Design}
\label{2.3}
Given the wide variety of neuron formats in QNN, a natural question arises: \textit{which QNN format is effective for CTR prediction tasks?} To answer this question, we observe from Table \ref{NF} that, without using Post Act, QNN consistently outperforms MLP, with several neuron formats achieving AUC levels of up to 81.50 (significantly surpassing MLP). Among these high-performing neuron formats, T19 achieves the best performance. Its structure is similar to T9, but the key difference is that T19 introduces \textbf{Mid Act}, which we define as activations applied within a single QNN layer before Hadmard Product. From the longer training epochs required for T19, we observe that T19 continues to improve as training progresses, whereas other QNN variants (e.g., CrossNetv2, T9, and T22) tend to overfit at earlier stages. Thus, we identify T19 as an effective neuron format and show that Mid Act enhances QNN's generalization. Moreover, T19 in Figure \ref{db} also exhibits a combination of sharp and smooth decision boundaries similar to DCNv2, achieving the highest accuracy. This indicates that introducing \textbf{Mid Act} into QNN enables it to integrate the sharp approximation capability brought by Act with its inherent smooth approximation capability, thereby achieving a complementary overall approximation.

Besides, several pairs of neuron formats are worth noting. For example, [T9 vs T23 and T19 vs T24] indicate that adding more parameters may not improve model performance, which is contrary to observations in computer vision \cite{starNet, resNet}. [T9 vs T18] show that using the sum operation to introduce residual connections is superior to the concat operation. [T1 vs T2 and T5 vs T23] demonstrate the importance of residual connections for QNN. [MLP, T1, T3, and T4] reveal that even the simple introduction of quadratic terms in traditional linear layers can reduce MLP's reliance on Act. [T1, T5, T9, T13, T16, and T24] are all representative baseline models designed by CTR researchers and can be regarded as specific types of QNN. Therefore, understanding feature interaction modeling in CTR from the perspective of QNN may provide new insights for future research. We recommend readers to refer to \cite{QNNbegin, qnn1, qnn2, qnn3, qnn4} for a broader and deeper understanding of QNN.

\section{A Better QNN Design for CTR prediction}
\begin{figure}[t]
    \centering
    \includegraphics[width=1\linewidth]{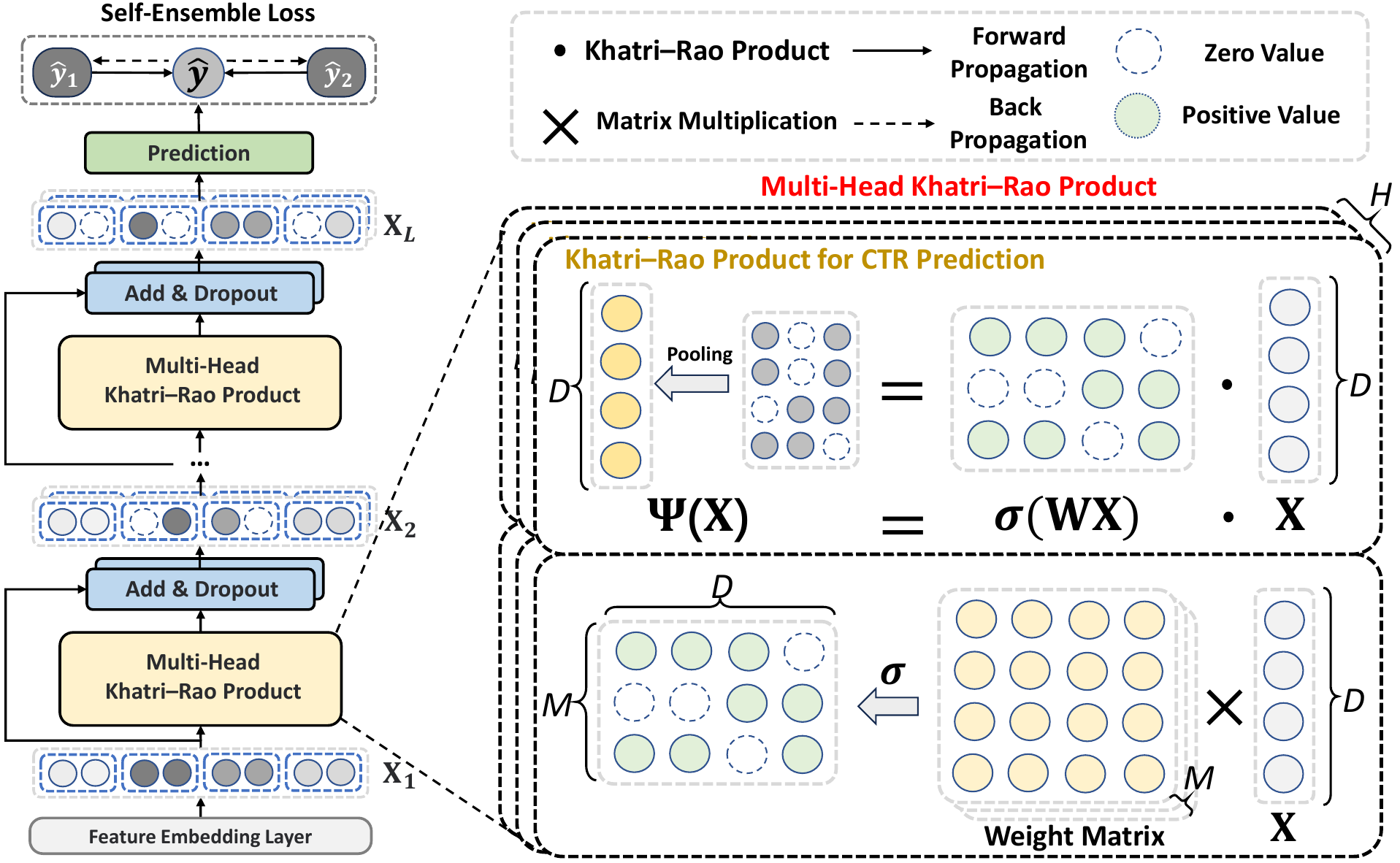}
    \captionsetup{justification=raggedright, font=small}
    \caption{Architecture of the QNN-$\alpha$.}
    \label{QNN}
    \vspace{-1em}
\end{figure}

Based on the conclusions in Section \ref{revis}, we believe that QNN holds significant potential for improving CTR model performance but remains underexplored. Therefore, in this section, we aim to further understand the underlying mechanisms of QNN neuron formats and make improvements to fully leverage the powerful approximation capabilities of QNN, thus proposing QNN-$\alpha$ as shown in Figure \ref{QNN}.

\subsection{Khatri–Rao Product: A Better Alternative to the Hadamard Product}
Hadamard Product (HP) has long been one of the most well-known methods for modeling feature interactions \cite{neuralvsmf, NCF, dcnv2, DCNv3}, demonstrating strong performance across various domains \cite{infinite, Quadralib, starNet} (as evidenced by the T19's superior performance). From the perspective of QNN, the effectiveness of HP lies in its ability to element-wisely expand the feature space and introduce nonlinear second-order polynomials $\chi$. For example, in Equation (\ref{CrossNetv2}), HP expands the feature space from 1 dimension to $D$ dimensions, while in Equation (\ref{ECN}), it extends it to $D^2$ dimensions. However, we argue that HP's lack of scalability limits the model capacity. To better understand this, we first revisit a simple HP:
\begin{equation}
\begin{aligned}
\mathbf{X}_a \odot \mathbf{X}_b=x_{a}^{i}x_{b}^{i}=\{x_a^1x_b^1, x_a^2x_b^2, \dots, x_a^Dx_b^D\}.
\end{aligned}
\end{equation}
It is evident that HP requires two matrices of the same shape, which forces $\mathbf{W}_l$ in Equation \ref{ECN} to always maintain a $D \times D$ scale, thereby reducing the network's scalability\footnote{ In this paper, we define scalability as the network's capability for horizontal scaling.}.

To address this issue, we initially attempt to introduce additional parameters or bottleneck structures to the neuron, as shown in [T9 vs T17 and T19 vs T25] in Table \ref{NF}. Unfortunately, this attempt reduces performance. we realize that this may be due to HP's overly simplistic method to handling information interactions in high-dimensional feature spaces, which fails to adapt to the diversity and complexity of feature distributions, thereby limiting the model's capacity. Subsequently, we find that the Khatri–Rao Product (KRP) \cite{Khatri-Rao} is a more expressive method for modeling feature interactions. Specifically, KRP performs a Kronecker Product \cite{KP} on each column of two feature matrices, generating a higher-dimensional interaction matrix. For easy understanding, here is a simple example:
\begin{equation}
\label{KRP}
\begin{aligned}
\mathbf{A}&=\left[\begin{array}{l|l}
\mathbf{A}^1 & \mathbf{A}^2 \end{array}\right] 
= \left[\begin{array}{l|l}
a^{1,1} & a^{1,2} \\
\end{array}\right], \\
\mathbf{B}&=\left[\begin{array}{l|l}
\mathbf{B}^1 & \mathbf{B}^2 \end{array}\right] 
= \left[\begin{array}{l|l}
b^{1,1} & b^{1,2} \\
b^{2,1} & b^{2,2}
\end{array}\right], \\
\mathbf{A} \bullet \mathbf{B} &=\left[\begin{array}{l|l}
\mathbf{A}^1 \otimes \mathbf{B}^1 & \mathbf{A}^2 \otimes \mathbf{B}^2
\end{array}\right]=\left[\begin{array}{l|l}
a^{1,1} b^{1,1} & a^{1,2} b^{1,2} \\
a^{1,1} b^{2,1} & a^{1,2} b^{2,2} \\
\end{array}\right] = \mathbf{C},
\end{aligned}
\end{equation}
where $\bullet$ denotes KRP, and $\otimes$ represents the Kronecker Product. From Equation (\ref{KRP}), we observe that when $\mathbf{B} = {b^{(1,:)}}$, KRP reduces to the HP. Therefore, we can consider KRP as a generalized version of HP. Furthermore, when $\mathbf{A} \in \mathbb{R}^D$ and $\mathbf{B} \in \mathbb{R}^{M \times D}$, the resulting $\mathbf{C} \in \mathbb{R}^{M \times D}$. However, in CTR prediction, following the CrossNetv2 paradigm, we aim for the neuron’s output to have the same dimensions as $\mathbf{X}$ to facilitate deeper computations and the introduction of residual connections. Therefore, dimensionality reduction of $\mathbf{C}$ is necessary, and sum pooling is a suitable choice for this purpose. Finally, we redefine KRP for CTR prediction as:
\begin{equation}
\label{CTRKRP}
\begin{aligned}
\mathbf{A} \bullet \mathbf{B} &=\left[\begin{array}{l|l}
\mathbf{A}^1 \otimes \mathbf{B}^1 & \mathbf{A}^2 \otimes \mathbf{B}^2
\end{array}\right]_{\text{sum}} \\
&=\left[\begin{array}{l|l}
a^{1,1} b^{1,1} + a^{1,1} b^{2,1} & a^{1,2} b^{1,2} + a^{1,2} b^{2,2} \\
\end{array}\right] = \mathbf{C}.
\end{aligned}
\end{equation}
If KRP is applied to T9, its neuron format is as follows:
\begin{equation}
\label{T26}
\begin{aligned}
\mathbf{X}_{l+1} &= \mathbf{X}_l \bullet \mathbf{W}_l \mathbf{X}_l,\ \ l=1,2,\dots,L, \\
x_{l+1}^i &\overset{l\ >\ 1}{=} x_{l-1}^i \sum_{j=1}^{D}\sum_{p=1}^{M} w_{l-1}^{p,i,j} x_{l-1}^j \sum_{k=1}^{D}\sum_{q=1}^{M} w_l^{q,i,k} x_l^k, \\
&= \underbrace{A_{(1,1)} x_{l-1}^1 x_l^1 + A_{(1,2)} x_{l-1}^1 x_l^2 + \cdots + A_{(D,D)} x_{l-1}^D x_l^D}_{D^2 \text { interaction items}},
\end{aligned}
\end{equation}
where $\mathbf{W}_l \in \mathbb{R}^{M \times D \times D}$, and $A_{(j, k)} = \sum_{p=1}^{M}\sum_{q=1}^{M} w_{l-1}^{p,i,j} w_{l}^{q,i,k} x_{l-1}^i$. In this way, we can set $M$ as a hyperparameter to adjust the model capacity and enhance scalability. 

\subsection{Multi-Head Khatri–Rao Product}
Theoretically, while KRP outperforms HP, it requires more computational resources, which is another issue we need to address. Considering the success of Transformers \cite{Attentionisallyouneed}, multi-head attention mechanisms have been widely used to reduce computational costs while achieving performance improvements. We incorporate this idea into KRP-based QNN and propose a new QNN-$\alpha$ version, whose neuron format is defined as follows:
\begin{equation}
\label{MHKRP}
\begin{aligned}
\Psi(\mathbb{X}^h) = \mathbb{X}^h \bullet \textbf{ReLU}(\mathbb{W}_\textit{a}^h \mathbf{X}) + \mathbb{X}^h, \\
\Phi(\mathbf{X}) = \Psi(\mathbb{X}^1)\ ||\ \Psi(\mathbb{X}^2)\ ||\ \dots\ ||\ \Psi(\mathbb{X}^H),
\end{aligned}
\end{equation}
where $\mathbb{X}^h \in \mathbb{R}^{\mathbb{D}}$ denotes the input of the $h$-th subspace, $\mathbb{W}_\textit{a} \in \mathbb{R}^{M \times \mathbb{D} \times D}$, $\mathbb{D} = D/H$, $H$ denotes the number of heads, and || denotes concatenation. This theoretically reduces both the parameter size and computational complexity of QNN-$\alpha$ by a factor of $H$, decreasing from $O(MD^2)$ to $O\left(MD^2/H\right)$.

\subsection{Self-Ensemble Loss}
While the Khatri-Rao Product effectively enhances the model's expressiveness and capacity, highly expressive models often exhibit instability when handling sparse data, increasing the risk of overfitting \cite{lightgcn}. To mitigate this, we draw inspiration from ensemble learning \cite{ensemble}, a widely adopted technique in CTR prediction \cite{deepfm, dcnv2, DCNv3, EKTF}. This method effectively reduces the prediction bias of a single network on specific samples by ensembling the predictions of multiple sub-networks, thereby enhancing overall generalization ability \cite{widedeep, deepfm}. However, directly relying on multiple sub-networks for ensembling not only increases computational and storage costs but also imposes higher resource demands for model deployment. To address this, inspired by the ideas of R-Drop \cite{R-drop} and Tri-BCE \cite{DCNv3}, we propose a \textbf{Self-Ensemble Loss (SE Loss)}, which eliminates the need for additional sub-networks. Instead, \textit{it dynamically forms an ensemble by aggregating multiple stochastic outputs from a single network during forward propagation, where the stochasticity is introduced by the use of dropout \cite{dropout}.} Meanwhile, it introduces a consistency constraint to reduce the discrepancy between the ensemble and individual forward results, thereby mitigating the prediction bias of the network. Let $\hat{y}_1 = \text{Forward}_1(X), \hat{y}_2 = \text{Forward}_2(X), \hat{y}=(\hat{y}_1 + \hat{y}_2)/2$, the loss is formulated in the following form:
\begin{equation}
\begin{aligned}
\label{SE Loss}
\mathcal{L}_{\scriptscriptstyle SE}=-\frac{1}{N} \sum_{i=1}^N\left[\tilde{y}_i \log \left(\hat{y}_{1,i}\hat{y}_{2,i}\right)+\left(1-\tilde{y}_i\right) \log \left(\left(1-\hat{y}_{1,i}\right)\left(1-\hat{y}_{2,i}\right)\right)\right],
\end{aligned}
\end{equation}
where $y$ denotes the true labels, $\tilde{y}$ represents the frozen $\hat{y}$, $N$ denotes the batch size, and the final loss is $\mathcal{L}_{\scriptscriptstyle total}=\mathcal{L}_{\scriptscriptstyle CTR} + \mathcal{L}_{\scriptscriptstyle SE}$, where $\mathcal{L}_{\scriptscriptstyle CTR}=-\frac{1}{N} \sum_{i=1}^N\left(y_i \log \left(\hat{y}_i\right)+\left(1-y_i\right) \log \left(1-\hat{y}_i\right)\right)$. Intuitively, $\mathcal{L}_{\scriptscriptstyle SE}$ forces the prediction results of two forward passes, $\hat{y}_{1,i}$ and $\hat{y}_{2,i}$, to simultaneously approach 1 or 0, thereby ensuring that the predictions of the two forwards remain consistent at the sample level. In this way, $\mathcal{L}_{\scriptscriptstyle SE}$ jointly constrains and optimizes the different prediction results of the single network, enabling mutual correction among the predictions. This reduces potential overfitting or bias issues in the network. Moreover, unlike the hard label supervision brought by $y$ in $\mathcal{L}_{\scriptscriptstyle CTR}$, $\mathcal{L}_{\scriptscriptstyle SE}$ introduces richer soft label information $\tilde{y}$ to the network. This allows the QNN to learn smoother decision boundaries during training, which is more conducive to handling complex distributions and challenging-to-classify samples \cite{labelsmooth}. Theoretically, we can perform more forward passes on the single network to obtain richer soft label information. However, this exponentially increases the training cost, which is unacceptable in industrial scenarios. Moreover, our experiments show that additional forward pass do not lead to significant performance improvements. Therefore, we recommend using only two forward passes. Notably, this idea can be understood or further improved from multiple perspectives, such as contrastive learning and knowledge distillation. We recommend referring to \cite{R-drop, mean-teacher, KDCL, SimCEN, mean-better} to help readers gain deeper technical insights.

\section{Experiments}
\label{Experiments}
In this section, we conduct comprehensive experiments on six CTR prediction datasets to validate the effectiveness, efficiency, scalability, and compatibility of our proposed QNN-$\alpha$.
\subsection{Experiment Setup}
\subsubsection{\textbf{Datasets.}} We evaluate QNN-$\alpha$ on six CTR prediction datasets: 
Tenrec\footnote{ \url{https://static.qblv.qq.com/qblv/h5/algo-frontend/tenrec_dataset.html}} \cite{Tenrec}, Criteo\footnote{ \url{https://www.kaggle.com/c/criteo-display-ad-challenge}} \cite{openbenchmark}, ML-1M\footnote{ \url{https://grouplens.org/datasets/movielens}} \cite{autoint}, Frappe\footnote{ \url{http://baltrunas.info/research-menu/frappe}} \cite{AFN, frappe}, iPinYou\footnote{ \url{https://contest.ipinyou.com/}} \cite{pnn2}, and KKBox\footnote{ \url{https://www.kkbox.com/intl}} \cite{Bars}. Table \ref{dataset}  provides detailed information about these datasets. A more detailed description of these datasets can be found in the given references and links. We follow the approach outlined in \cite{openbenchmark}. For the Criteo dataset, we discretize the numerical feature fields by rounding down each numeric value $x$ to $\lfloor \log^2(x) \rfloor$ if $x > 2$, and $x = 1$ otherwise. We set a threshold to replace infrequent categorical features with a default "OOV" token. We set the threshold to 10 for Criteo, KKBox, and Tenrec, 2 for iPinYou, and 1 for the small dataset ML-1M and Frappe. More specific data processing procedures and results can be found in our run logs\footnote{ \url{https://github.com/salmon1802/QNN/tree/main/checkpoints} \label{footnote:checkpoint}}.
\begin{table}[t]
\tiny
\renewcommand\arraystretch{1}
\centering
\captionsetup{font=small}
\caption{Dataset statistics}
\label{dataset}
\resizebox{0.9\linewidth}{!}{
\begin{tabular}{cccc} 
\hline
\textbf{Dataset} & \textbf{\#Instances} & \textbf{\#Fields} & \textbf{\#Features} \\
\hline 
\textbf{Tenrec} & 120,342,306 & 15  & 3,404,996 \\
\textbf{Criteo} & 45,840,617 & 39  & 910,747 \\
\textbf{ML-1M} & 739,012  & 7  & 9,751 \\
\textbf{Frappe} & 288,609  & 10  & 5,383 \\
\textbf{iPinYou} & 19,495,974 & 16  & 665,765\\
\textbf{KKBox} & 7,377,418 & 13  & 91,756\\ \hline
\end{tabular}}
\vspace{-1em}
\end{table}
\subsubsection{\textbf{Evaluation Metrics.}} To compare the performance, we utilize two commonly used metrics in CTR models: \textbf{Logloss}, \textbf{AUC} \cite{SimCEN, RFM, ComboFashion}. AUC stands for Area Under the ROC Curve, which measures the probability that a positive instance will be ranked higher than a randomly chosen negative one. Logloss is the result of the calculation of $\mathcal{L}_{CTR}$. A lower Logloss suggests a better capacity for fitting the data. Besides, \textbf{Latency} is an important evaluation metric in CTR prediction. Therefore, we calculate the average inference latency per 100 samples on the test set.

\begin{table*}[t]
\renewcommand\arraystretch{1}
\centering
\captionsetup{font=footnotesize}
\caption{Performance comparison of different deep CTR models. "*": Integrating the original model with DNN networks. Meanwhile, we conduct a two-tailed T-test to assess the statistical significance between our models and the best baseline ($\star$: $p$ < 1e-3). \textit{Abs.Imp} represents the absolute performance improvement of QNN over the strongest baseline. Typically, CTR researchers consider an improvement of \textit{0.001 (0.1\%)} in Logloss and AUC to be statistically significant \cite{dcn,EDCN,CL4CTR,openbenchmark}.} 
\label{baselines}
\resizebox{\linewidth}{!}{
\begin{tabular}{ccccccccccccc}
\hline
  \multicolumn{1}{c|}{} &
  \multicolumn{2}{c|}{\textbf{Tenrec}} &
  \multicolumn{2}{c|}{\textbf{Criteo}} &
  \multicolumn{2}{c|}{\textbf{ML-1M}} &
  \multicolumn{2}{c|}{\textbf{Frappe}} &
  \multicolumn{2}{c|}{\textbf{iPinYou}} &
  \multicolumn{2}{c}{\textbf{KKBox}}\\ \cline{2-13} 
  \multicolumn{1}{c|}{\multirow{-2}{*}{\textbf{Models}}} &
  \multicolumn{1}{c}{Logloss$\downarrow$} &
  \multicolumn{1}{c|}{AUC(\%)$\uparrow$} &
  \multicolumn{1}{c}{Logloss$\downarrow$} &
  \multicolumn{1}{c|}{AUC(\%)$\uparrow$} & 
  \multicolumn{1}{c}{Logloss$\downarrow$} &
  \multicolumn{1}{c|}{AUC(\%)$\uparrow$} &
  \multicolumn{1}{c}{Logloss$\downarrow$} &
  \multicolumn{1}{c|}{AUC(\%)$\uparrow$} &
  \multicolumn{1}{c}{Logloss$\downarrow$} &
  \multicolumn{1}{c|}{AUC(\%)$\uparrow$} &
  \multicolumn{1}{c}{Logloss$\downarrow$} &
  \multicolumn{1}{c}{AUC(\%)$\uparrow$} \\
  \hline
  \multicolumn{1}{c|}{DNN \cite{DNN}} &
  0.4366 &
  \multicolumn{1}{c|}{80.21} &
  0.4380 &
  \multicolumn{1}{c|}{81.40} & 
  0.3100 &
  \multicolumn{1}{c|}{90.30} &
  0.1653 &
  \multicolumn{1}{c|}{98.11} &
   0.005545 &
  \multicolumn{1}{c|}{78.06} &
  0.4811 &
  \multicolumn{1}{c}{85.01} \\
  \multicolumn{1}{c|}{PNN \cite{pnn1}} &
  0.4358 &
  \multicolumn{1}{c|}{80.28} &
  0.4378 &
  \multicolumn{1}{c|}{81.42} &
  0.3070 &
  \multicolumn{1}{c|}{90.42} &
  0.1405 &
  \multicolumn{1}{c|}{98.45} &
  0.005544 &
  \multicolumn{1}{c|}{78.13} &
  0.4793 &
  \multicolumn{1}{c}{85.15} \\
  \multicolumn{1}{c|}{Wide \& Deep \cite{widedeep}} &
  0.4366 &
  \multicolumn{1}{c|}{80.22} &
  0.4376 &
  \multicolumn{1}{c|}{81.42} &
  0.3056 &
  \multicolumn{1}{c|}{90.45} &
  0.1525 &
  \multicolumn{1}{c|}{98.32} &
  0.005542 &
  \multicolumn{1}{c|}{78.09} &
  0.4852 &
  \multicolumn{1}{c}{85.04} \\
  \multicolumn{1}{c|}{DeepFM \cite{deepfm}} &
  0.4365 &
  \multicolumn{1}{c|}{80.23} &
  0.4375 &
  \multicolumn{1}{c|}{81.43} &
  0.3073 &
  \multicolumn{1}{c|}{90.51} &
  0.1575 &
  \multicolumn{1}{c|}{98.37} &
  0.005549 &
  \multicolumn{1}{c|}{77.94} &
  0.4785 &
  \multicolumn{1}{c}{85.31}\\
  \multicolumn{1}{c|}{DCNv1 \cite{dcn}} &
  0.4360 &
  \multicolumn{1}{c|}{80.26} &
  0.4376 &
  \multicolumn{1}{c|}{81.44} &
  0.3156 &
  \multicolumn{1}{c|}{90.38} &
  0.1544 &
  \multicolumn{1}{c|}{98.38} &
  0.005541 &
  \multicolumn{1}{c|}{78.13} &
  \underline{0.4766} &
  \multicolumn{1}{c}{85.31}\\
  \multicolumn{1}{c|}{xDeepFM \cite{xdeepfm}} &
   0.4363 &
  \multicolumn{1}{c|}{80.24} &
  0.4376 &
  \multicolumn{1}{c|}{81.43} &
  0.3054 &
  \multicolumn{1}{c|}{90.47} &
  0.1509 &
  \multicolumn{1}{c|}{98.45} &
  0.005534 &
  \multicolumn{1}{c|}{78.25} &
  0.4772 &
  \multicolumn{1}{c}{85.35} \\
  \multicolumn{1}{c|}{AutoInt* \cite{autoint}} &
  0.4362 &
  \multicolumn{1}{c|}{80.25} &
  0.4390 &
  \multicolumn{1}{c|}{81.32} &
  0.3112 &
  \multicolumn{1}{c|}{90.45} &
  0.1520 &
  \multicolumn{1}{c|}{98.41} &
  0.005544 &
  \multicolumn{1}{c|}{78.16} &
  0.4773 &
  \multicolumn{1}{c}{85.34} \\
  \multicolumn{1}{c|}{AFN* \cite{AFN}} &
  0.4357 &
  \multicolumn{1}{c|}{80.30} &
  0.4384 &
  \multicolumn{1}{c|}{81.38} &
  0.3048 &
  \multicolumn{1}{c|}{90.53} &
   0.1598 &
  \multicolumn{1}{c|}{98.19} &
  0.005539 &
  \multicolumn{1}{c|}{78.17} &
  0.4842 &
  \multicolumn{1}{c}{84.89}\\
  \multicolumn{1}{c|}{DCNv2 \cite{dcnv2}} &
  0.4359 &
  \multicolumn{1}{c|}{80.28} &
  0.4376 &
  \multicolumn{1}{c|}{81.45} &
  0.3098 &
  \multicolumn{1}{c|}{90.56} &
  0.1484 &
  \multicolumn{1}{c|}{98.45} &
  0.005539 &
  \multicolumn{1}{c|}{78.26} &
  0.4787 &
  \multicolumn{1}{c}{85.31}\\
  \multicolumn{1}{c|}{EDCN \cite{EDCN}} &
  0.4356 &
  \multicolumn{1}{c|}{80.30} &
  0.4386 &
  \multicolumn{1}{c|}{81.36} &
  0.3073 &
  \multicolumn{1}{c|}{90.48} & 
  0.1620 &
  \multicolumn{1}{c|}{98.41} &
  0.005573 &
  \multicolumn{1}{c|}{77.93} & 
  0.4952 &
  \multicolumn{1}{c}{85.27} \\
  \multicolumn{1}{c|}{MaskNet \cite{masknet}} &
  0.4363 &
  \multicolumn{1}{c|}{80.23} &
  0.4387 &
  \multicolumn{1}{c|}{81.34} &
  0.3080 &
  \multicolumn{1}{c|}{90.34} &
  0.1916 &
  \multicolumn{1}{c|}{98.32} &
  0.005556 &
  \multicolumn{1}{c|}{77.85} &
  0.5003 &
  \multicolumn{1}{c}{84.79}\\
  \multicolumn{1}{c|}{CL4CTR \cite{CL4CTR}} &
  0.4363 &
  \multicolumn{1}{c|}{80.24} &
  0.4383 &
  \multicolumn{1}{c|}{81.35} &
  0.3074 &
  \multicolumn{1}{c|}{90.33} &
  0.1559 &
  \multicolumn{1}{c|}{98.27} &
  0.005543 &
  \multicolumn{1}{c|}{78.06} &
  0.4972  &
  \multicolumn{1}{c}{83.78}\\
  \multicolumn{1}{c|}{EulerNet \cite{EulerNet}} &
  0.4362 &
  \multicolumn{1}{c|}{80.24} &
  0.4379 &
  \multicolumn{1}{c|}{81.47} &
  0.3050 &
  \multicolumn{1}{c|}{90.44} &
  \underline{0.1331}  &
  \multicolumn{1}{c|}{98.50} &
  0.005540 &
  \multicolumn{1}{c|}{78.30} &
  0.4922 &
  \multicolumn{1}{c}{84.27}\\ 
  \multicolumn{1}{c|}{FinalMLP \cite{finalmlp}} &
  0.4359 &
  \multicolumn{1}{c|}{80.28} &
  0.4373 &
  \multicolumn{1}{c|}{81.45} &
  0.3058 &
  \multicolumn{1}{c|}{90.52} &
  0.1472 &
  \multicolumn{1}{c|}{\underline{98.51}} &
  0.005556 &
  \multicolumn{1}{c|}{78.02} &
  0.4822 &
  \multicolumn{1}{c}{85.10}\\ 
  \multicolumn{1}{c|}{FINAL(2B) \cite{FINAL}} &
  \underline{0.4355} &
  \multicolumn{1}{c|}{\underline{80.31}} &
  0.4371 &
  \multicolumn{1}{c|}{81.49} &
  0.3035 &
  \multicolumn{1}{c|}{90.53} &
  0.1392 &
  \multicolumn{1}{c|}{98.47} &
  0.005540 &
  \multicolumn{1}{c|}{78.13} &
  0.4800 &
  \multicolumn{1}{c}{85.14} \\
  \multicolumn{1}{c|}{RFM \cite{RFM}} &
  0.4358 &
  \multicolumn{1}{c|}{80.29} &
  0.4374 &
  \multicolumn{1}{c|}{81.47} &
  0.3048 &
  \multicolumn{1}{c|}{90.51} &
  0.1476 &
  \multicolumn{1}{c|}{98.48} &
  0.005540 &
  \multicolumn{1}{c|}{78.25} &
  0.4853 &
  \multicolumn{1}{c}{84.70}\\
  \multicolumn{1}{c|}{ECN \cite{DCNv3}} &
  0.4361 &
  \multicolumn{1}{c|}{80.23} &
  \underline{0.4364} &
  \multicolumn{1}{c|}{\underline{81.55}} &
  \underline{0.3013} &
  \multicolumn{1}{c|}{\underline{90.59}} &
  0.1587 &
  \multicolumn{1}{c|}{98.49} &
  \underline{0.005534} &
  \multicolumn{1}{c|}{\underline{78.43}} &
  0.4778 &
  \multicolumn{1}{c}{\underline{85.40}}\\ \hline
  \multicolumn{1}{c|}{\textbf{QNN-$\alpha$}} &
  \textbf{0.4349$^\star$} &
  \multicolumn{1}{c|}{\textbf{80.35$^\star$}} &
  \textbf{0.4358$^\star$} &
  \multicolumn{1}{c|}{\textbf{81.63$^\star$}} &
   \textbf{0.2960$^\star$} &
  \multicolumn{1}{c|}{\textbf{90.87$^\star$}} &
   \textbf{0.1313$^\star$} &
  \multicolumn{1}{c|}{\textbf{98.62$^\star$}} &
  \textbf{0.005516$^\star$} &
  \multicolumn{1}{c|}{\textbf{78.63$^\star$}} &
  \textbf{0.4730$^\star$} &
  \multicolumn{1}{c}{\textbf{85.76$^\star$}} \\ \hline
   \multicolumn{1}{c|}{\textit{Abs.Imp}} &
  -0.0006 &
  \multicolumn{1}{c|}{+0.04} &
  -0.0006 &
  \multicolumn{1}{c|}{+0.08} &
   -0.0053 &
  \multicolumn{1}{c|}{+0.28} &
   -0.0018 &
  \multicolumn{1}{c|}{+0.11} &
  -0.000018 &
  \multicolumn{1}{c|}{+0.20} &
  -0.0036 &
  \multicolumn{1}{c}{+0.36}\\
  \hline
\end{tabular}}
\end{table*}

\subsubsection{\textbf{Baselines.}}
\label{section_baseline}
We compared QNN-$\alpha$ with some SOTA models. Further, we select several high-performance CTR representative baselines, such as PNN \cite{pnn1} and Wide \& Deep \cite{widedeep} (2016); DeepFM \cite{deepfm} and DCNv1 \cite{dcn} (2017); xDeepFM (2018) \cite{xdeepfm} (2018); AutoInt* (2019) \cite{autoint}; AFN* (2020) \cite{AFN}; DCNv2 \cite{dcnv2} and EDCN \cite{EDCN}, MaskNet \cite{masknet} (2021); CL4CTR \cite{CL4CTR}, EulerNet \cite{EulerNet}, FinalMLP \cite{finalmlp}, FINAL \cite{FINAL} (2023); RFM \cite{RFM}, ECN \cite{DCNv3} (2024). 

\subsubsection{\textbf{Implementation Details.}} We implement all models using PyTorch \cite{PYTORCH} and refer to existing works \cite{openbenchmark, FuxiCTR}. We employ the Adam optimizer \cite{adam} to optimize all models, with a default learning rate set to 0.001. For the sake of fair comparison, we set the embedding dimension to 128 for KKBox and 16 for the other datasets \cite{openbenchmark, Bars}. The batch size is set to 4,096 on the ML-1M, Frappe, and iPinYou datasets and 10,000 on the other datasets. To prevent overfitting, we employ early stopping with a patience value of 2. During training, we employ a Reduce-LR-on-Plateau scheduler that reduces the learning rate by a factor of 10 when performance stops improving in any given epoch \cite{openbenchmark,Bars}. The hyperparameters of the baseline model are configured and fine-tuned based on the \textit{optimal values} provided in  \cite{FuxiCTR,openbenchmark} and their original paper. All experiments in this paper are conducted under the same environment using NVIDIA GeForce RTX 4090 GPU. Further details on model hyperparameters and dataset configurations are available in our straightforward and accessible running logs\footref{footnote:checkpoint}.

\subsection{Overall Performance}
To validate the effectiveness of QNN-$\alpha$, we conduct a performance comparison across six datasets. The experimental results are presented in Table \ref{baselines}, where bold numbers indicate the best performance, and underlined numbers represent the second-best performance. We can draw the following conclusions:
\begin{itemize}[leftmargin=*]
\item Ensemble models consistently achieve better performance. For example, DCNv2, FinalMLP, and FINAL(2B) show significant performance improvements compared to a standalone DNN.
\item QNN-$\alpha$ achieves a new SOTA across all datasets with only a single network.
\item Models categorized as QNN, like FINAL(2B) and ECN, show strong performance, ranking second on Tenrec, Criteo, ML-1M, iPinYou, and KKBox, validating their effectiveness.
\item After thorough tuning, the performance differences between models on the Tenrec and Criteo datasets become minimal, which is also reported in \cite{openbenchmark}. This highlights the difficulty of achieving performance gains on large-scale sparse datasets. Meanwhile, minor gains on open datasets may yield significant improvements in industrial applications \cite{finalmlp}.
\end{itemize}

\subsection{In-Depth Study of QNN-$\alpha$}
\subsubsection{\textbf{Ablation Study.}}
To investigate the impact of each component of QNN-$\alpha$ on its performance, we conduct experiments on several variants of QNN-$\alpha$: (1) \underline{w/o Res}: removes the residual term in Equation (\ref{MHKRP}), (2) \underline{w/o Mid Act}: removes the activation function before the KRP operation in Equation (\ref{MHKRP}), (3) \underline{w/o KRP}: replaces the KRP in Equation (\ref{MHKRP}) with a linear transformation, (4) \underline{w/o SE Loss}: removes the Self-Ensemble Loss.

\begin{table}[t]
\Huge
\renewcommand\arraystretch{1.2}
\centering
\captionsetup{font=small}
\caption{\textbf{Ablation study of QNN-$\alpha$.}} 
\resizebox{\linewidth}{!}{
\begin{tabular}{c|cc|cc|cc|cc}
\hline
\multirow{2}{*}{\textbf{Model}}  & \multicolumn{2}{c|}{\textbf{Criteo}} & \multicolumn{2}{c|}{\textbf{ML-1M}} & \multicolumn{2}{c|}{\textbf{Frappe}} & \multicolumn{2}{c}{\textbf{KKBox}} \\ \cline{2-9} 
                       & $ \text{Logloss}\downarrow$       & $ \text{AUC(\%)}\uparrow$        & $ \text{Logloss}\downarrow$       & $ \text{AUC(\%)}\uparrow$   &
                       $ \text{Logloss}\downarrow$       & $ \text{AUC(\%)}\uparrow$ &
                       $ \text{Logloss}\downarrow$       & $ \text{AUC(\%)}\uparrow$ \\ \hline
w/o Res                & \textbackslash{} & \textbackslash{} & 0.2971       & 90.78       & 0.1468        & 98.51       & \textbackslash{} & \textbackslash{} \\
w/o Mid Act & 0.4362  & 81.57 & 0.2969  & 90.79 & 0.1453  & 98.49 & 0.4773  & 85.57 \\
w/o KRP     & 0.4380   & 81.38 & 0.3016  & 90.66 & 0.1630   & 98.51 & 0.4901  & 84.66 \\
w/o SE Loss & 0.4363  & 81.57 & 0.2970   & 90.78 & 0.1581  & 98.45 & 0.4783  & 85.57 \\ \hline
QNN-$\alpha$        & 0.4358  & 81.63 & 0.2960   & 90.87 & 0.1313  & 98.62 & 0.4730   & 85.76 \\ \hline
\end{tabular}}
\label{ablation}
\end{table}

Table \ref{ablation}\footnote{ "\textbackslash" indicates that the network fails to train success.}, we observe that all variants of QNN experience some degree of performance degradation, which indicates the necessity of each component. w/o Res highlights the importance of the residual term for QNN. Removing it leads to gradient explosion on some datasets, making the model unable to train properly. w/o Mid Act shows that while post-activation (Post Act) reduces QNN's performance, placing the activation function in an appropriate position further improves the network's performance. w/o KRP contributes the most to performance loss, with a degradation of over 1\% on the KKBox dataset, demonstrating the effectiveness of KRP. w/o SE Loss confirms that the two forward passes and the consistency constraints contribute to the model's performance improvements.

\subsubsection{\textbf{Impact of Different Network Depths in QNN-$\alpha$}}
\begin{figure}[t]
 
    \setlength{\belowcaptionskip}{0pt}
    \centering
    \begin{minipage}[t]{0.33\linewidth}
        \centering
        \includegraphics[width=\linewidth]{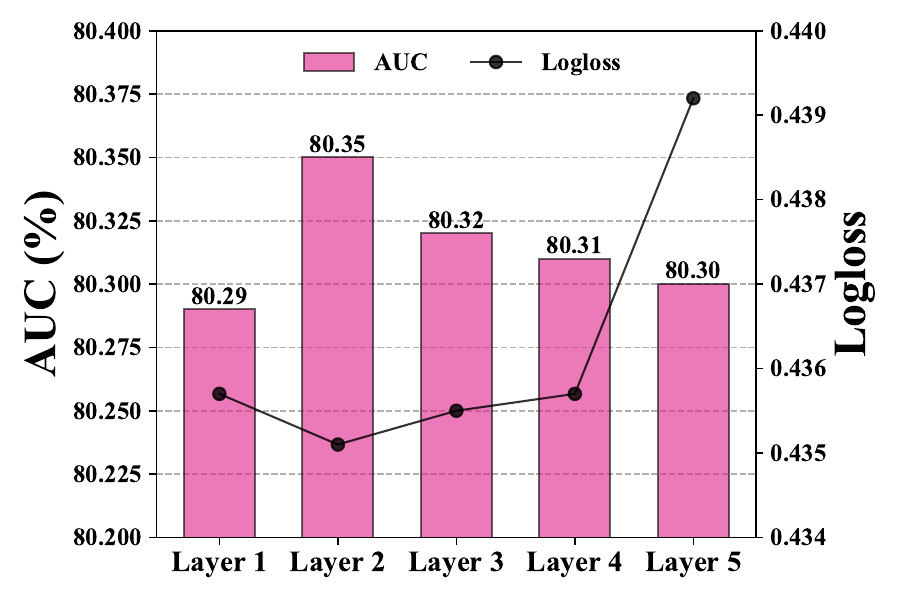}
         \subcaptionsetup{font=footnotesize}
        \subcaption{Tenrec}
    \end{minipage}%
    \begin{minipage}[t]{0.33\linewidth}
        \centering
        \includegraphics[width=\linewidth]{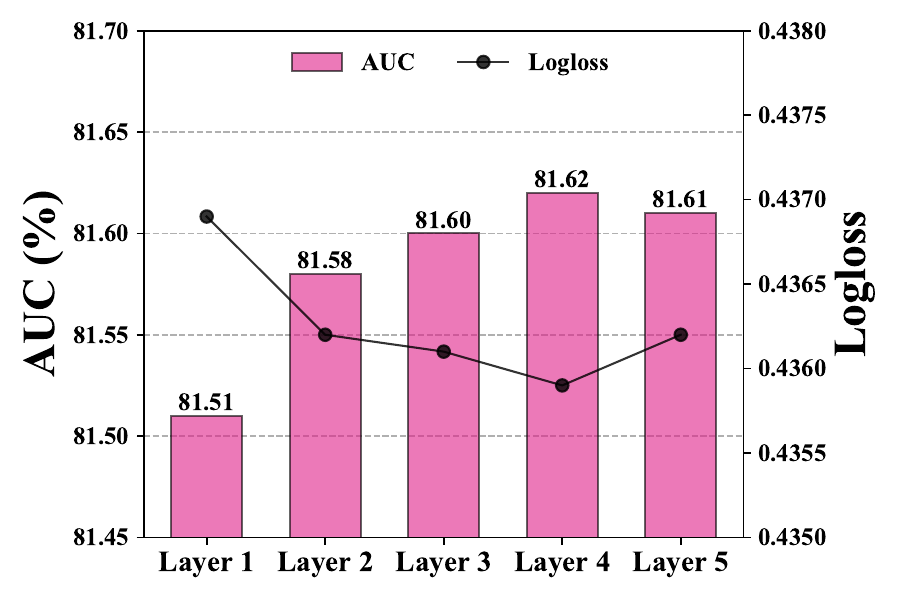}
        \subcaptionsetup{font=footnotesize}
        \subcaption{Criteo}
    \end{minipage}%
    \begin{minipage}[t]{0.33\linewidth}
        \centering
        \includegraphics[width=\linewidth]{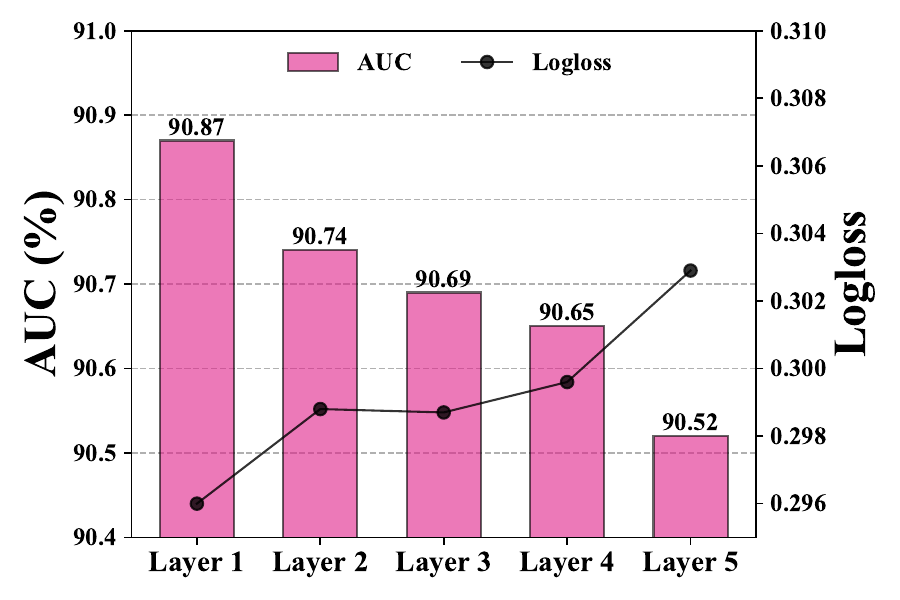}
        \subcaptionsetup{font=footnotesize}
        \subcaption{ML-1M}
    \end{minipage}
    \captionsetup{justification=raggedright, font=small}
    \caption{Performance of different $L$ in QNN-$\alpha$.}
    \label{Layer_T26}
\end{figure}

The depth of a neural network is often an important hyperparameter \cite{resNet}. To investigate the impact of network depth $L$ on the performance of QNN-$\alpha$, we conduct experiments with different depths on the Tenrec, Criteo, and ML-1M datasets. As shown in Figure \ref{Layer_T26}, QNN-$\alpha$ achieves optimal performance with 2-layer, 4-layer, and 1-layer settings on the Tenrec, Criteo, and ML-1M datasets, respectively, followed by a performance decline. This indicates that networks that are either too deep or too shallow can result in performance degradation. We recommend setting the search space for $L$ to \{1, 2, 3, 4\} during hyperparameter tuning as a good practice.

\subsubsection{\textbf{Impact of Different Mid Act in QNN-$\alpha$}}
\begin{figure}[t]
    \centering
    \begin{minipage}[t]{0.33\linewidth}
        \centering
        \includegraphics[width=\linewidth]{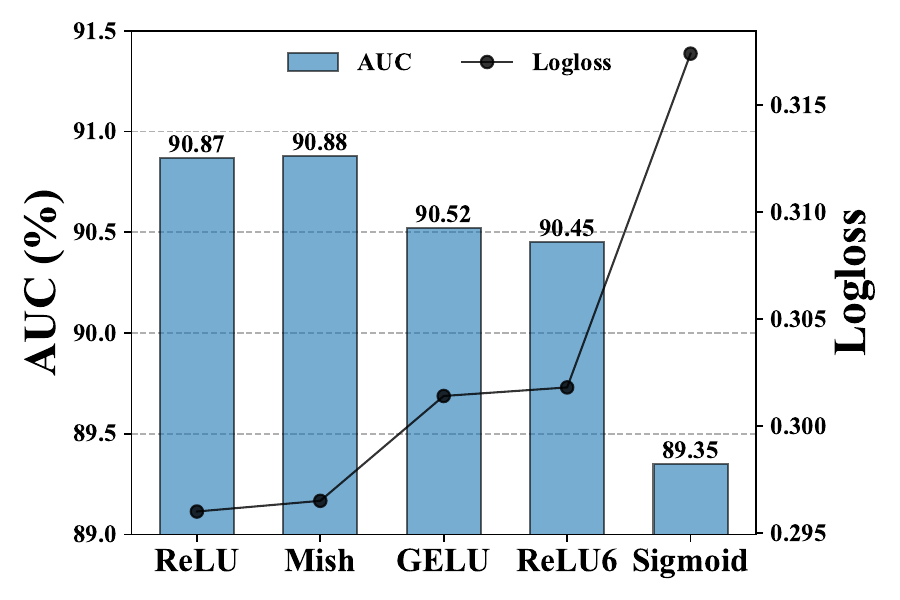}
        \subcaptionsetup{font=footnotesize}
        \subcaption{ML-1M}
    \end{minipage}%
    \begin{minipage}[t]{0.33\linewidth}
        \centering
        \includegraphics[width=\linewidth]{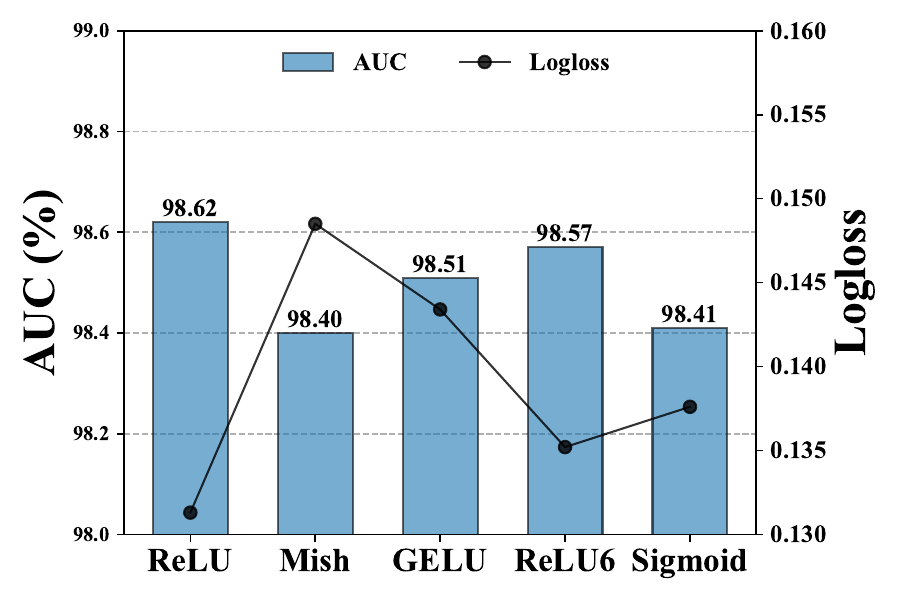}
        \subcaptionsetup{font=footnotesize}
        \subcaption{Frappe}
    \end{minipage}%
    \begin{minipage}[t]{0.33\linewidth}
        \centering
        \includegraphics[width=\linewidth]{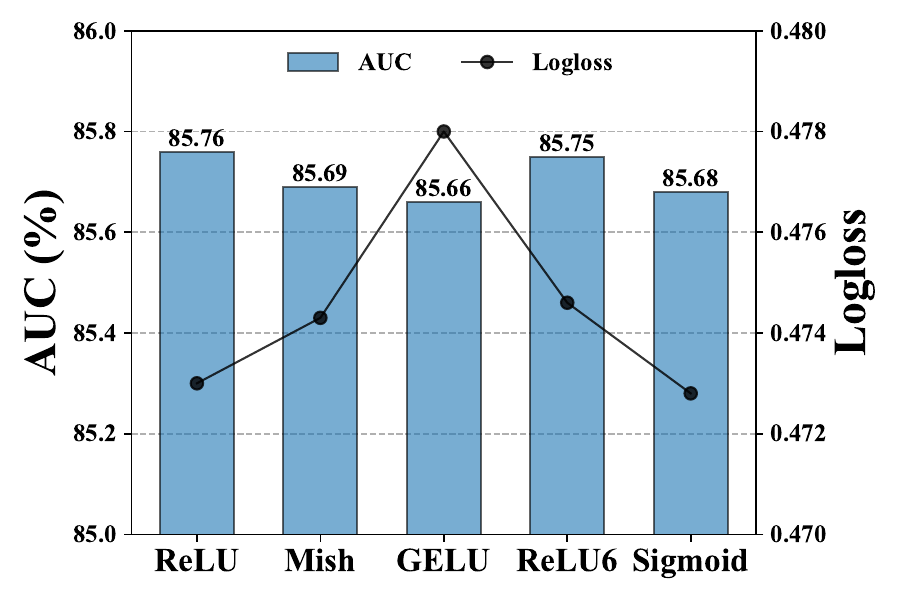}
        \subcaptionsetup{font=footnotesize}
        \subcaption{KKBox}
    \end{minipage}
    \captionsetup{justification=raggedright, font=small}
    \caption{Performance of different Mid Act in QNN-$\alpha$.}
    \label{Act_T26}
\end{figure}

In the design of our QNN-$\alpha$, we follow most CTR models \cite{FINAL, dcnv2, xdeepfm}, where the ReLU activation function is used as the Mid Act. To verify the reasonableness of this design, we further experiment with various activation functions. The experimental results, shown in Figure \ref{Act_T26}, indicate that ReLU demonstrates competitive performance across multiple datasets. This suggests that using a more complex Act does not provide additional performance improvements for QNN, especially those without a zero-value region (e.g., Sigmoid). 


\subsubsection{\textbf{Impact of Different Loss in QNN-$\alpha$}}
\begin{figure}[t]
    \centering
    \begin{minipage}[t]{0.33\linewidth}
        \centering
        \includegraphics[width=\linewidth]{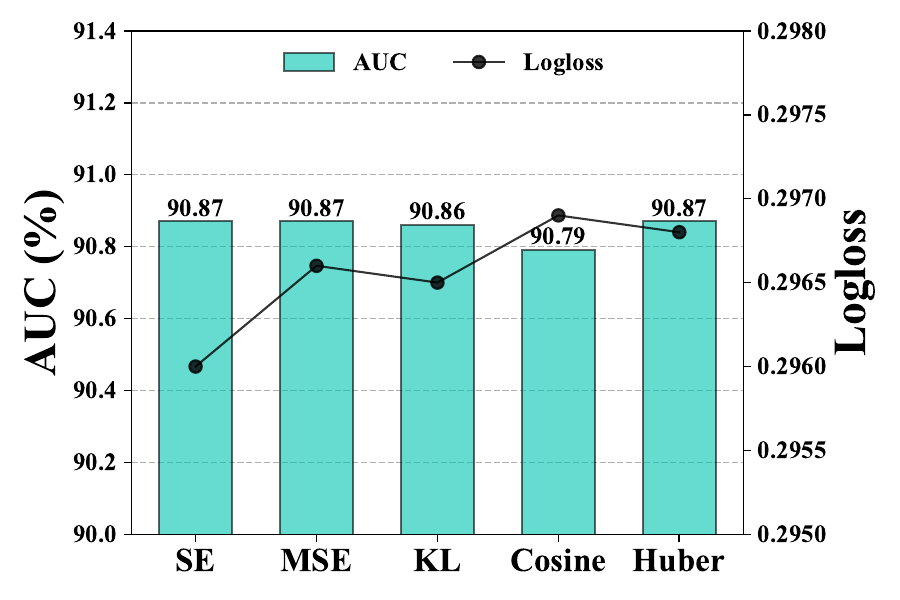}
        \subcaptionsetup{font=footnotesize}
        \subcaption{ML-1M}
    \end{minipage}%
    \begin{minipage}[t]{0.33\linewidth}
        \centering
        \includegraphics[width=\linewidth]{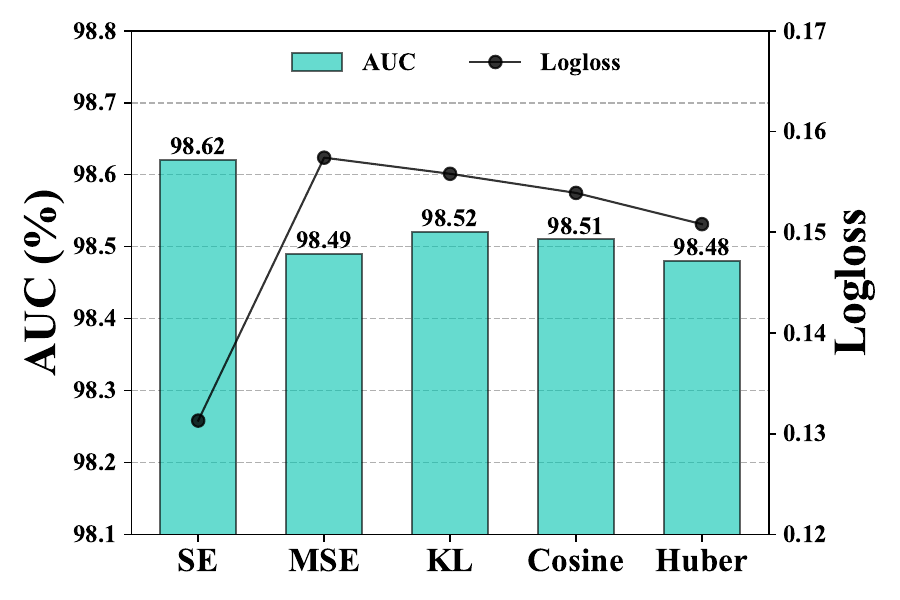}
        \subcaptionsetup{font=footnotesize}
        \subcaption{Frappe}
    \end{minipage}%
    \begin{minipage}[t]{0.33\linewidth}
        \centering
        \includegraphics[width=\linewidth]{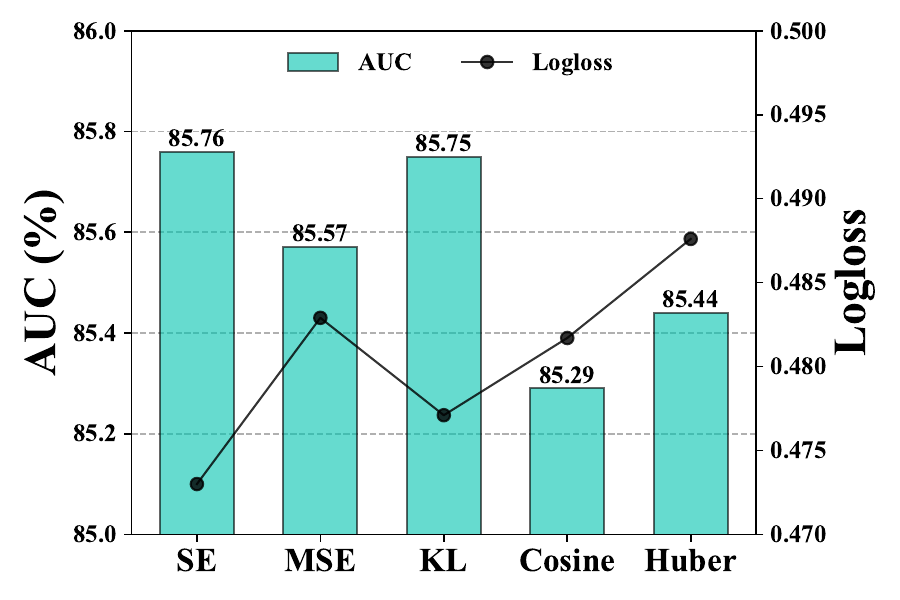}
        \subcaptionsetup{font=footnotesize}
        \subcaption{KKBox}
    \end{minipage}
    \captionsetup{justification=raggedright, font=small}
    \caption{Performance of different Loss in QNN-$\alpha$.}
    \label{Loss_T26}
\end{figure}

To validate the effectiveness of the proposed SE Loss, we compare it with several commonly used consistency losses. Specifically, we replace only the loss function without altering the computation process of $\hat{y}, \hat{y}_1$, and $\hat{y}_2$. The experimental results, shown in Figure \ref{Loss_T26}, demonstrate that SE Loss achieves the best performance across all three datasets. For example, on the ML-1M dataset, while AUC values for the five losses are similar, SE Loss achieves the lowest Logloss, demonstrating its effectiveness in improving classification capability \cite{rankandlog}. On the KKBox dataset, SE Loss also outperforms KL Loss, further supporting this conclusion. On the Frappe dataset, SE Loss shows a clear advantage, with both AUC and Logloss significantly better than other loss functions, including a notable Logloss reduction at the 0.01 level. These results underscore the strong generalization and optimization performance of SE Loss. Additionally, we recommend setting the dropout rate to 0.1, as it works well across all datasets in our experiments, while higher rates consistently degrade performance.

\subsubsection{\textbf{Scalability Study}}
\begin{figure}[t]

    \centering
    \begin{minipage}[t]{0.32\linewidth}
        \centering
        \includegraphics[width=\linewidth]{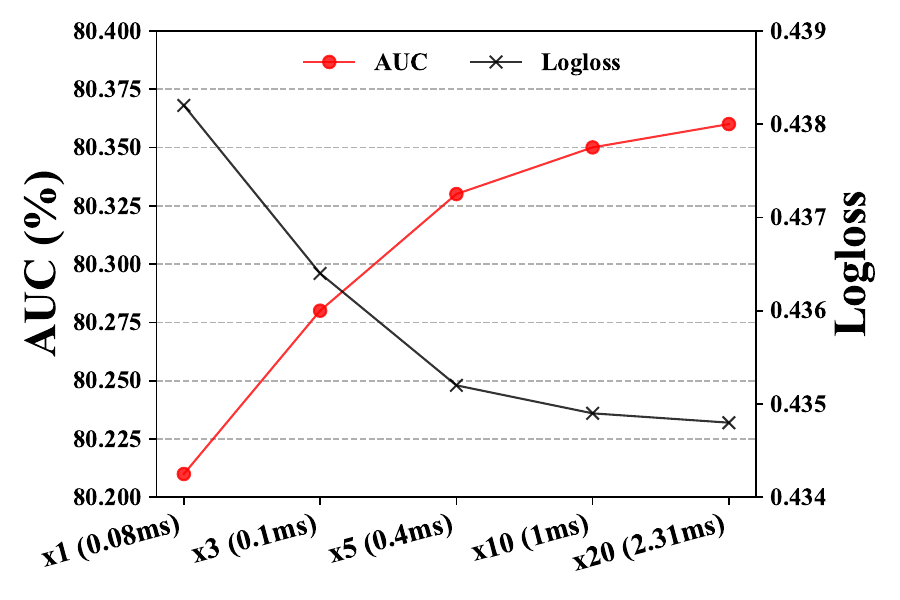}
        \subcaptionsetup{font=footnotesize}
        \subcaption{Tenrec}
    \end{minipage}
    \begin{minipage}[t]{0.32\linewidth}
        \centering
        \includegraphics[width=\linewidth]{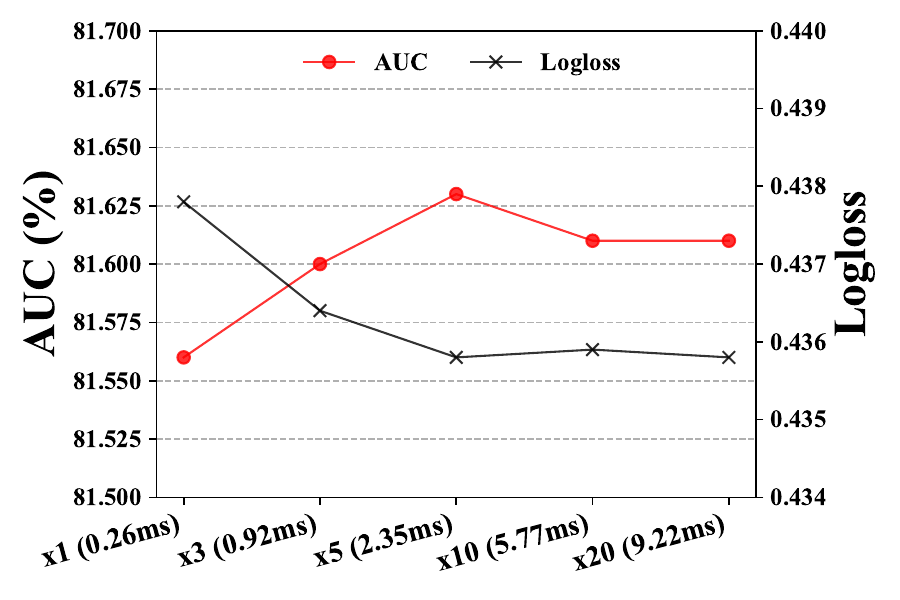}
        \subcaptionsetup{font=footnotesize}
        \subcaption{Criteo}
    \end{minipage}
    \begin{minipage}[t]{0.32\linewidth}
        \centering
        \includegraphics[width=\linewidth]{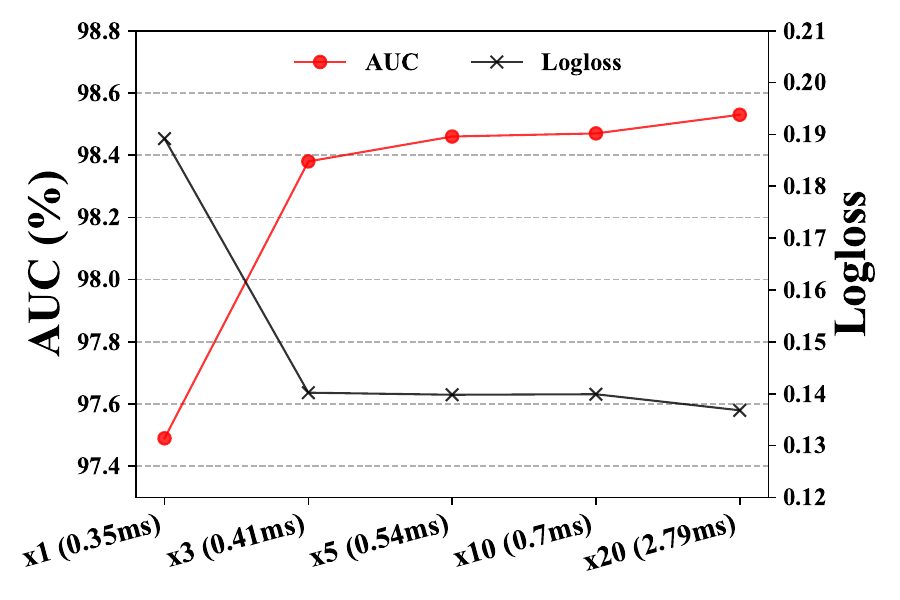}
        \subcaptionsetup{font=footnotesize}
        \subcaption{Frappe}
    \end{minipage}%
    \captionsetup{justification=raggedright, font=small}
    \caption{Performance of different $M$ in QNN-$\alpha$.}
    \label{M_T26}
\end{figure}

Scalability has always been one of the key concerns for deep learning models \cite{scaling}. A good CTR model should flexibly adjust its capacity to adapt to different data patterns based on varying application scenarios. Theoretically, the hyperparameter $M$ in the KRP of Equation (\ref{T26}) effectively meets this requirement. To validate the scalability of KRP, we conduct experiments on two large-scale, highly sparse datasets and a small dataset. Figure \ref{M_T26} shows that as $M$ increases, Logloss decreases and AUC increases on the Tenrec and Frappe datasets, demonstrating the scalability of QNN-$\alpha$. Notably, when $M=1$, the model performs worst as the KRP in Equation (\ref{T26}) degenerates into a simple HP. This validates the superiority of our proposed KRP.

\begin{table}[t]
\Huge
\renewcommand\arraystretch{1.2}
\centering
\captionsetup{font=small}
\caption{\textbf{Compatibility study of QNN-$\alpha$.}} 
\resizebox{\linewidth}{!}{
\begin{tabular}{c|cc|cc|cc}
\hline
\multirow{2}{*}{\textbf{Model}}  & \multicolumn{2}{c|}{\textbf{ML-1M}} & \multicolumn{2}{c|}{\textbf{Frappe}} & \multicolumn{2}{c}{\textbf{iPinYou}}\\ \cline{2-7} 
                       & $ \text{Logloss}\downarrow$       & $ \text{AUC(\%)}\uparrow$        & $ \text{Logloss}\downarrow$       & $ \text{AUC(\%)}\uparrow$   &
                       $ \text{Logloss}\downarrow$       & $ \text{AUC(\%)}\uparrow$ \\ \hline
\multicolumn{1}{c|}{DNN \cite{DNN}}  & 0.3100 & 90.30 & 0.1653 & 98.11  & 0.005545 & 78.06     \\
\multicolumn{1}{c|}{DNN + SE Loss}  & 0.3050 & 90.52 & 0.1616 & 98.25  & 0.005543 & 78.08     \\ 
\multicolumn{1}{c|}{QNN-$\alpha$}  & \textbf{0.2960} & \textbf{90.87} & \textbf{0.1313} & \textbf{98.62}  & \textbf{0.005516} & \textbf{78.63} \\ \hline
\multicolumn{1}{c|}{xDeepFM \cite{xdeepfm}}  & 0.3070 & 90.42 & 0.1405 & 98.45  & 0.005544 & 78.13     \\
\multicolumn{1}{c|}{xDeepFM + SE Loss}  & 0.3042 & 90.50 & 0.1363 & 98.52  & 0.005544 & 78.15     \\ 
\multicolumn{1}{c|}{xDeepFM (QNN-$\alpha$)}  & \textbf{0.2957} & \textbf{90.86} & \textbf{0.1329} & \textbf{98.61} & \textbf{0.005524} & \textbf{78.44}     \\ \hline
\multicolumn{1}{c|}{DeepFM \cite{deepfm}}   & 0.3073 & 90.51 & 0.1575 & 98.37   & 0.005549  & 77.94    \\
\multicolumn{1}{c|}{DeepFM + SE Loss}  & 0.3055 & 90.58 & 0.1504 & 98.53  & 0.005548 & 77.97     \\ 
\multicolumn{1}{c|}{DeepFM (QNN-$\alpha$)}  & \textbf{0.2966} & \textbf{90.82} & \textbf{0.1364} & \textbf{98.60} & \textbf{0.005533} & \textbf{78.27}     \\ \hline
\multicolumn{1}{c|}{DCNv2 \cite{dcnv2}}  & 0.3070 & 90.42 & 0.1484 & 98.45  & 0.005539 & 78.26     \\
\multicolumn{1}{c|}{DCNv2 + SE Loss}  & 0.3037 & 90.53 & 0.1449 & 98.56  & 0.005536 & 78.20     \\ 
\multicolumn{1}{c|}{DCNv2 (QNN-$\alpha$)}  & \textbf{0.2957} & \textbf{90.91} & \textbf{0.1380} & \textbf{98.63}  & \textbf{0.005527} & \textbf{78.49}     \\ \hline
\end{tabular}}
\label{Compatibility}
\end{table}

\subsubsection{\textbf{Compatibility Study}}
QNN-$\alpha$ can be viewed as a plug-and-play module designed to enhance the performance of deep CTR prediction models. To validate the compatibility of QNN-$\alpha$, we replace the original MLP in four representative baseline models with QNN-$\alpha$. The experimental results, shown in Table \ref{Compatibility}, indicate that the QNN-enhanced baseline models consistently achieve performance improvements and outperform their variants that only use SE Loss. This demonstrates the advantages of QNN in terms of both performance and compatibility. From an industrial application perspective, as increasingly complex ensemble models continue to emerge \cite{CTRensemble1, PINTEREST}, QNN-$\alpha$ can serve as a flexible component within sophisticated architectures, further boosting overall performance.

\subsubsection{\textbf{Inference Efficiency Study}}
Inference efficiency is crucial in industrial recommender systems \cite{openbenchmark}. High inference latency can limit the practical use of accurate models. To evaluate QNN-$\alpha$, we compare it with FinalMLP \cite{finalmlp} and FINAL \cite{FINAL}, which have been successfully applied in production environments. While KRP-based feature interaction increases computational cost, our multi-head KRP method alleviates this issue effectively. Experiments shown in Table \ref{head} explore the impact of different head numbers ($H$) on performance and latency. Results show that increasing $H$ reduces inference latency, confirming multi-head KRP's in optimizing inference efficiency. When $H=4$, $H=4$, and $H=2$, QNN-$\alpha$ achieves optimal performance on ML-1M, iPinYou, and KKBox datasets, respectively. It surpasses FINAL in prediction accuracy and significantly reduces latency, demonstrating its strong performance and efficiency for real-world production environments.

\begin{table}[t]
\Huge
\renewcommand\arraystretch{1.2}
\centering
\captionsetup{font=small}
\caption{Inference efficiency study of different $H$ in QNN-$\alpha$.}
\resizebox{\linewidth}{!}{
\begin{tabular}{cc|ccc|ccc|ccc}
\hline
\multicolumn{2}{c|}{\multirow{2}{*}{\textbf{Model}}} &
  \multicolumn{3}{c|}{\textbf{ML-1M}} &
  \multicolumn{3}{c|}{\textbf{iPinYou}} &
  \multicolumn{3}{c}{\textbf{KKBox}} \\ \cline{3-11} 
\multicolumn{2}{c|}{} &
  Logloss &
  AUC(\%) &
  Latency &
  Logloss &
  AUC(\%) &
  Latency &
  Logloss &
  AUC(\%) &
  Latency \\ \hline
  \multicolumn{2}{c|}{FinalMLP \cite{finalmlp}} &
  0.3058 &
  90.52 &
  1.10ms &
  0.005556 &
  78.02 &
  0.78ms &
  0.4822 &
  85.10 &
  7.62ms \\
\multicolumn{2}{c|}{FINAL (2B) \cite{FINAL}} &
  0.3035 &
  90.53 &
  0.68ms &
  0.005540 &
  78.13 &
  0.44ms &
  0.4795 &
  85.08 &
  8.65ms \\ \hline
\multicolumn{1}{c|}{\multirow{4}{*}{\begin{tabular}[c]{@{}c@{}}QNN\\-$\alpha$\end{tabular}}} &
  x1 Head &
  0.2965 &
  90.82 &
  0.27ms &
  0.005536 &
  78.13 &
  0.79ms &
  \textbf{0.4726} &
  85.74 &
  8.69ms \\
\multicolumn{1}{c|}{} &
  x2 Head &
  \textbf{0.2960} &
  90.87 &
  0.26ms &
  0.005529 &
  78.22 &
  0.51ms &
  0.4730 &
  \textbf{85.76} &
  7.52ms \\
\multicolumn{1}{c|}{} &
  x4 Head &
  \textbf{0.2960} &
  \textbf{90.88} &
  0.24ms &
  \textbf{0.005516} &
  \textbf{78.63} &
  0.23ms &
  0.4763 &
  85.73 &
  7.13ms \\
\multicolumn{1}{c|}{} &
  x8 Head &
  0.2997 &
  90.62 &
  \textbf{0.18ms} &
  0.005529 &
  78.28 &
  \textbf{0.21ms} &
  0.4807 &
  85.66 &
  \textbf{5.75ms} \\ \hline
\end{tabular}}
\label{head}
\end{table}

\subsubsection{\textbf{Training Efficiency Study}}
\begin{table}[t]
\Huge
\renewcommand\arraystretch{1.2}
\centering
\captionsetup{font=small}
\caption{Training efficiency study of QNN-$\alpha$.}
\resizebox{\linewidth}{!}{
\begin{tabular}{c|cccc|cccc}
\hline
                                 & \multicolumn{4}{c|}{\textbf{Criteo}} & \multicolumn{4}{c}{\textbf{KKBox}} \\ \cline{2-9} 
\multirow{-2}{*}{\textbf{Model}} & Logloss  & AUC(\%)    & Memory   & Time & Logloss  & AUC(\%)   & Memory  & Time \\ \hline
xDeepFM &
0.4376 &
  81.43 &
  15806MiB &
  781s &
  0.4772 &
  85.35 &
  11799MiB &
  109s \\
AutoInt*                         & 0.4390   & 81.32  & 5158MiB  & 392s  & 0.4773   & 85.34 & 6449MiB & 56s   \\ \hline
\textbf{w/o SE Loss} &
  0.4363 &
  81.57 &
  \textbf{3996MiB} &
  \textbf{258s} &
  0.4783 &
  85.57 &
  \textbf{5795MiB} &
  \textbf{39s} \\
\textbf{QNN-$\alpha$}                   & \textbf{0.4358}   & \textbf{81.63}  & 5304MiB  & 328s  & \textbf{0.4730}   & \textbf{85.76} & 8227MiB & 63s   \\ \hline
\end{tabular}}
\label{Training_efficiency}
\vspace{-1em}
\end{table}

Training efficiency is also a major concern for CTR researchers~\cite{openbenchmark,DCNv3,GDCN}. In real-world production environments, users generate data on the scale of hundreds of millions per day. Low training efficiency increases the additional costs of recommender systems. To evaluate QNN-$\alpha$, we compare its training efficiency with xDeepFM~\cite{xdeepfm} and AutoInt*~\cite{autoint}. The experimental results are shown in Table~\ref{Training_efficiency}. We observe that QNN-$\alpha$ achieves GPU memory requirements and per-epoch training time similar to AutoInt*, while delivering significant performance improvements. Meanwhile, since SE Loss only requires two forward passes through QNN without additional operations on the embedding layer, its introduction does not result in a twofold decrease in training efficiency. In practice, researchers can decide whether to incorporate SE Loss according to specific requirements.

\section{Related Work}
\subsection{CTR Prediction}
Effectively capturing feature interactions is one of the key factors in improving the performance of CTR prediction models \cite{widedeep, dcn, deepfm, OptFusion}. Early CTR models are typically limited to capturing low-order feature interactions with bounded degrees \cite{FM, FMFM, AFM, FwFM} or rely on hand-crafted feature combinations based on expert knowledge \cite{widedeep, LR}. With the rise of deep learning, some works attempt to leverage MLP to implicitly capture high-order feature interactions \cite{NFM, deepfm, pnn1, pnn2, dcn, DIN}, achieving promising results. However, research suggests that while MLP can theoretically serve as universal function approximators \cite{MLPapproximators}, they struggle to effectively model product-based feature interactions \cite{neuralvsmf}. 

As a result, CTR researchers begin to explore paradigms that combine explicit and implicit feature interactions \cite{dcnv2, xdeepfm, masknet}. In recent years, increasingly sophisticated explicit feature interaction methods have been proposed \cite{fignn, FGCNN, PEPNET, adagin, AFN, autoint}. To evaluate these methods, CTR researchers establish open benchmarking systems \cite{openbenchmark}, which reveal that, under well-tuned hyperparameter settings, the performance differences between models are often minimal. Consequently, some works attempt to break away from this conventional modeling paradigm and provide novel insights. Examples include only MLP-based \cite{finalmlp}, complex-space \cite{EulerNet, RFM}, exponential \cite{DCNv3, FINAL}, contrastive learning-based \cite{CL4CTR, SimCEN, MISS}, and LLM-enhanced \cite{EASE, LLMsCTR-multidomain} feature interaction models.

\subsection{Quadratic Neural Networks} 
QNN is a class of neural architectures designed to explicitly model second-order interactions between input features \cite{qnn1}. Unlike traditional MLP to learn implicitly, QNN leverages quadratic terms to capture feature interactions more directly and expressively \cite{qnn2, qnn3, qnn4}. Moreover, some studies demonstrate that QNN has stronger nonlinear approximation capabilities compared to MLP \cite{QNNniu}, and QNN can also be regarded as a universal function approximator \cite{QNNbegin}. QuadraLib \cite{Quadralib} explores the optimization and design of QNN architectures. More recently, several works focus on further improving the computational efficiency of QNN \cite{QuadraNet, QuadraNetv2} and exploring higher-dimensional neural networks \cite{infinite}. To the best of our knowledge, no existing work has attempted to think about the CTR prediction task from the perspective of QNN, even though classic models such as FM \cite{FM} and CrossNetv2 \cite{dcnv2} can be directly interpreted as a QNN. Therefore, this work has the potential to provide new insights into CTR prediction.

\section{Conclusion}
In this paper, we revisited feature interaction models for CTR prediction from the QNN perspective. We found that the linearly independent quadratic polynomials in QNN were the core reason why HP played a significant role, as they expanded the feature space and introduced smooth nonlinear approximation capabilities. We evaluated 25 QNN neuron formats and observed that the traditional Post Act did not effectively enhance QNN's performance. We then proposed QNN-$\alpha$, incorporating the Multi-Head Khatri-Rao Product as an improved HP and a Self-Ensemble Loss for dynamic ensemble predictions without extra sub-networks. Experiments show that QNN-$\alpha$ achieves a new SOTA performance on six public datasets while ensuring low latency, scalability, and compatibility. 

Besides, it is noteworthy that we further integrate QNN with user behaviors sequence-based CTR prediction models and propose the Quadratic Interest Network~\cite{QIN}. This method achieves an impressive second place in the EReL@MIR~\cite{EReL@MIR} competition, further demonstrating the effectiveness and practical value of our method.


\begin{acks}
This work is supported by the National Science Foundation of China (No. 62272001 and No. 62206002).
\end{acks}

\bibliographystyle{ACM-Reference-Format}
\balance
\bibliography{sample-base}


\end{document}